\documentclass[prd,twocolumn,superscriptaddress,floatfix,nofootinbib]{revtex4-2}
\pdfoutput=1 
\usepackage[T1]{fontenc} 
\usepackage{amsmath,amssymb,amsfonts,amsthm}
\usepackage{graphicx}
\usepackage[usenames,dvipsnames]{color}
\usepackage{enumitem}
\usepackage{verbatim}
\usepackage[percent]{overpic}
\usepackage{rotating}
\usepackage{hyperref}
\usepackage{array}
\usepackage{mathtools}
\usepackage{lmodern}
\usepackage[normalem]{ulem}
\usepackage{braket}
\usepackage{tensor}
\usepackage{dsfont}
\usepackage{bm}
\usepackage{tikz}

\usepackage{mathrsfs}

\usetikzlibrary{decorations.pathmorphing}
\usepackage{CJKutf8}

\newcommand{\kgy}[1]{{\color{blue}\bf [Ken: #1]}}

\usetikzlibrary{positioning}
\usetikzlibrary{calc,through,backgrounds}

\theoremstyle{definition}
\newtheorem{definition}{Definition}[section]
\newtheorem{theorem}{Theorem}[section]

\newcommand{\kako}[1]{\left( #1 \right)}
\newcommand{\kagikako}[1]{\left[ #1 \right]}
\newcommand{\ts}[1]{ _{\text{#1}} }
\newcommand{\Bigkako}[1]{\Big( #1 \Big)}

\newcommand{\Bigkagikako}[1]{\Big[ #1 \Big]}

\newcommand{\erf}{\text{erf}}
\newcommand{\erfc}{\text{erfc}}
\newcommand{\erfi}{\text{erfi}}
\newcommand{\hatb}[1]{\hat{\bm{#1}}}

\allowdisplaybreaks

\DeclareMathOperator{\Tr}{Tr}
\newcommand{\R}{\mathbb{R}}
\newcommand{\dd}{\text{d}}
\newcommand{\bk}{{\bm{k}}}
\newcommand{\bx}{{\bm{x}}}

\newcommand{\id}{\mathds{1}}
\newcommand{\sx}{\mathsf{x}}

\newcommand{\ii}{\mathsf{i}}

\newcommand{\kk}{|\bm{k}|}

\newcommand{\hil}{\mathcal{H}}

\begin{document}

\title{Decoherence of spin superposition state caused by a quantum electromagnetic field}


\author{Kensuke Gallock-Yoshimura}
\email{gallockyoshimura.kensuke@phys.kyushu-u.ac.jp} 

\affiliation{Department of Physics, Kyushu University, 744 Motooka, Nishi-Ku, Fukuoka 819-0395, Japan}

\author{Yuuki Sugiyama}
 \email{sugiyama.yuki@issp.u-tokyo.ac.jp}
\affiliation{Institute for Solid State Physics, the University of Tokyo, Kashiwa, Chiba 277-8581, Japan}

\author{Akira Matsumura}
\email[]{matsumura.akira@phys.kyushu-u.ac.jp}
\affiliation{Department of Physics, Kyushu University, 744 Motooka, Nishi-Ku, Fukuoka 819-0395, Japan}

\author{Kazuhiro Yamamoto}
\email{yamamoto@phys.kyushu-u.ac.jp}
\affiliation{Department of Physics, Kyushu University, 744 Motooka, Nishi-Ku, Fukuoka 819-0395, Japan}
\affiliation{Research Center for Advanced Particle Physics, Kyushu University,
744 Motooka, Nishi-ku, Fukuoka 819-0395, Japan}
\affiliation{International Center for Quantum-Field Measurement Systems for Studies of the Universe
and Particles (QUP), KEK, Oho 1-1, Tsukuba, Ibaraki 305-0801, Japan}

\begin{abstract}
In this study, we investigate the decoherence of a spatially superposed electrically neutral spin-$\frac12$ particle in the presence of a relativistic quantum electromagnetic field in Minkowski spacetime. 
We demonstrate that decoherence due to the spin-magnetic field coupling can be categorized into two distinct factors: local decoherence, originating from the two-point correlation functions along each branch of the superposed trajectories, and nonlocal decoherence, which arises from the correlation functions between the two superposed trajectories. 
These effects are linked to phase damping and amplitude damping. 
We also show that if the quantum field is prepared in a thermal state, decoherence monotonically increases with the field temperature. 
\end{abstract}

\maketitle
\flushbottom

\section{Introduction}


In recent years, the quantum superposition of the center-of-mass (COM) degree of freedom of a particle has attracted considerable attention, particularly in studies exploring the intersection between quantum theory and gravity. 
Seminal works of Bose \textit{et al.} \cite{Bose:2017nin} and Marletto and Vedral \cite{Marletto:2017kzi} proposed an experimental scheme to test the quantumness of gravity, known as the Bose-Marletto-Vedral (BMV) experiment---also referred to as the quantum gravity induced entanglement of masses (QGEM) protocol. 
In this protocol, two spatially superposed masses experiencing the Newtonian gravitational potential are expected to generate entanglement if gravity is a quantum entity.

These proposals employ a Stern-Gerlach-type scheme to generate spatial superposition states. 
Let $\ket{C}$ be a mass' COM state. 
The mass is assumed to have an embedded spin, which is initialized in a superposition state, $\ket{\uparrow}+\ket{\downarrow}$ up to a normalization constant. 
It is then exposed to an inhomogeneous classical magnetic field, and the mass becomes spatially superposed as $\ket{\downarrow, C_\downarrow} + \ket{\uparrow, C_\uparrow}$, where $\ket{C_\downarrow}$ and $\ket{C_\uparrow}$ denote the trajectories corresponding to the spin state (e.g., left and right trajectories). 
The BMV experiment employs two copies of such a spatial superposition state to test the quantum nature of gravity. 

A primary focus of spatial superposition states, in both experimental and theoretical contexts, is decoherence. 
In the BMV experiment, decoherence hinders our ability to detect quantum gravity-induced entanglement. 
As a result, developing methods to mitigate these decoherence effects is of paramount importance. 
Related studies have been carried out such as the effect of decoherence on the gravitationally induced entanglement \cite{Chevalier.Witnessing.2020, Rijavec.Decoherence.2021, Sugiyama:2022ixw, Hidaka:2022gsv, Gunnink.Gravitational.2023} and methods for overcoming decoherence \cite{Schut.Improving.2022, Tilly.Qudits.2021, Pedernales.Motional.2020}.

From a theoretical perspective, the decoherence of spatially superposed particles is also crucial. 
A thought experiment, originally proposed in \cite{mari2016experiments}, identifies an inconsistency between the principles of causality and complementarity. 
It suggests that an observer's decision can affect the coherence of a spatially superposed mass, even if they are causally disconnected. 
This problem was further analyzed and resolved in \cite{Belenchia.superposition.2018, Belenchia.superposition.2019, Danielson.newton.2022} by considering the decoherence of spatially superposed particles and vacuum fluctuations. 
See \cite{Sugiyama:2022wcd, Hidaka:2022tzk, Sugiyama:2023tqu} for further analyses on the effect of decoherence on this thought experiment, and \cite{Danielson.BH.decohere.2022, Danielson.Killing.decohere.2023, danielson2024LocalDescription} in the context of black hole spacetime.

These studies on the decoherence of spatially superposed particles typically focus on the COM degree of freedom. 
However, to our knowledge, little investigation has been done on the decoherence of a spatially superposed \textit{spin degree of freedom} caused by spin-magnetic field coupling. 
This is particularly interesting because the embedded spin in a mass is responsible for generating spatial superposition, yet it cannot avoid the decoherence from the vacuum fluctuations of a quantum electromagnetic field even if the mass is electrically neutral. 
See \cite{Ford.electron.coherence.1993, Ford.electron.coherence.1997, Breuer.coherence.2001, Mazzitelli.Decoherence.2003} for the decoherence of an electron coupled to a quantum electromagnetic field, and \cite{Sharifian:2023jem} for gravity-induced decoherence on a spin-$\frac12$ system.

In this paper, we analyze the decoherence of a spin's superposition state caused by a quantum electromagnetic field. 
Assuming the spin particle, which is a nonrelativistic first quantized system, is prepared in a spatial superposition state, we allow the superposed spin particle to interact with a relativistic quantum electromagnetic field via spin-magnetic field coupling for a finite period. 
For simplicity, we consider a superposed trajectory at rest (i.e., no acceleration or relative velocity) in $(3+1)$-dimensional Minkowski spacetime. 
Nevertheless, our analysis can be readily extended to a spin particle traveling along arbitrary trajectories and to quantum electromagnetic fields in an $(n+1)$-dimensional curved spacetime. 
We carry out a perturbative analysis to obtain the final density matrix for the spin particle and compute a decoherence measure. 
We find that the spatially superposed spin particle undergoes decoherence due to amplitude damping and phase damping effects. 
Each of these effects contributes to the decoherence measure both locally and nonlocally. 
The local contribution depends solely on each branch of the superposed trajectory and is independent of the spatial splitting configuration. 
Conversely, the nonlocal contribution depends on the field correlation functions between the two branches of the superposed trajectory. 
Depending on the spatial superposition configuration, the nonlocal contribution may mitigate the decoherence caused by the local terms. 
Finally, we consider the quantum electromagnetic field in a thermal state and show that the spin particle monotonically decoheres with the field temperature.

The paper is organized as follows. 
In Sec.~\ref{sec:setup}, we review the quantum theory of electromagnetic fields and the Hamiltonian describing the interaction between a single spin and the field. 
Assuming a small coupling constant, we employ a perturbative analysis.
In Sec.~\ref{sec:final density matrix}, we derive the final density matrix for the spin system, which is initially entangled with the COM degree of freedom. 
We provide concrete expressions of the elements of the final density matrix when the field is initially in the Minkowski vacuum and thermal states. 
We then introduce the $l_1$ norm of coherence as a coherence measure and define the measure for decoherence in Sec.~\ref{sec:Decoherence of spin}. 
The decoherence effect due to vacuum fluctuations and thermal noise is examined. 
Our conclusion is given in Sec.~\ref{sec:conclusion}. 
Appendix~\ref{app:elements in coherence} provides detailed calculations for the elements of the final density matrix, and Appendix~\ref{app:UDW} discusses the case of the scalar field.

Unless otherwise stated, we work in units where $\hbar = c= k\ts{B}= 1$ and use the mostly-plus signature convention, $(-,+,+,+)$.

\section{Setup}\label{sec:setup}

\subsection{Quantum electromagnetic field}\label{subsec:QED}
We begin by reviewing the quantized electromagnetic field. 
Consider a classical Lagrangian density for the electromagnetic field in $(3+1)$-dimensional Minkowski spacetime, 
\begin{align}
    \mathcal{L}
    &=
        -\dfrac{1}{4} F_{\mu \nu} F^{\mu \nu}\,,
\end{align}
where $F_{\mu \nu}= \partial_\mu A_\nu - \partial_\nu A_\mu$ is the field strength tensor with $A^\mu$ being the four-potential. 
In the radiation gauge $A^0=0$, $\bm{\nabla} \cdot \bm{A}=0$, the equation of motion becomes 
\begin{align}
    \square \bm A(\sx)=0\,,
\end{align}
where $\square$ is the d'Alembert operator and $\sx$ is a spacetime point.

Quantizing the field in the Heisenberg picture, the vector potential operator $\hatb{A}(\sx)$ can be written as a mode decomposition: 
\begin{align}
    &\hat{\bm{A}}(\sx)= \int_{\R^3} 
        \dfrac{ \dd^3k }{ \sqrt{(2\pi)^3 2\kk } }
        \sum_{\lambda=1}^2
        \kako{
            \bm{e}_{(\bk,\lambda)} e^{ -\ii \kk t + \ii \bk \cdot \bx } \hat a_{\bk, \lambda}
            + 
            \text{h.c.}
        },
\end{align}
where $\bm{e}_{(\bk,\lambda)}$ is the polarization vector with $\lambda$ representing the two kinds of polarizations, and it obeys 
\begin{subequations}
    \begin{align}
        &\bm{e}_{(\bk,\lambda)} \cdot \bm{e}_{(\bk,\lambda')}^*= \delta_{\lambda, \lambda'}\,, \\
        &\bk \cdot \bm{e}_{(\bk,\lambda)} =0 \quad \forall \lambda \,.
    \end{align}
\end{subequations}
The Minkowski vacuum $\ket{0}$ is defined to be 
\begin{align}
    \hat a_{\bk, \lambda} \ket{0}=0
    \quad 
    \forall \bk, \lambda \,,
\end{align}
and the commutation relation for the creation and annihilation operators reads 
\begin{align}
    [\hat a_{\bk, \lambda}, \hat a_{\bk', \lambda'}^\dag]
    &=
        \delta^{(3)} (\bk - \bk') \delta_{\lambda, \lambda'}\,.
\end{align}
The quantum magnetic field $\hatb{B}(\sx) \coloneqq \bm{\nabla} \times \hatb{A}(\sx)$ reads 
\begin{align}
    \hatb{B}(\sx)
    &=
        \int_{\R^3} 
        \dd^3 k 
        \sum_{\lambda=1}^2 
        \kako{
            \bm{B}_{(\bk, \lambda)}(\sx) \hat a_{\bk, \lambda}
            + 
            \bm{B}_{(\bk, \lambda)}^*(\sx) \hat a_{\bk, \lambda}^\dag
        }\,, \label{eq:B mode expansion}
\end{align}
where 
\begin{align}
    \bm{B}_{(\bk, \lambda)}(\sx)
    &\equiv 
        \dfrac{ \ii \bk \times \bm{e}_{(\bk,\lambda)} }{ \sqrt{ (2\pi)^3 2 \kk } }
        e^{ -\ii \kk t + \ii \bk \cdot \bx } \label{eq:magnetic mode function}
\end{align}
is the mode function for the magnetic field. 
In this paper, we will use the following projector: 
\begin{align}
    \sum_\lambda \bm{e}_{(\bk,\lambda)}^{(i)} \bm{e}_{(\bk,\lambda)}^{(j)*}
    =
        \delta^{ij} - \dfrac{k^i k^j }{ \kk^2 }\,, \label{eq:projector}
\end{align}
where $\bm{e}_{(\bk,\lambda)}^{(i)}$ and $k^i$ are the components of the vectors $\bm{e}_{(\bk,\lambda)}$ and $\bk$, respectively.

We comment on reparametrization of the time variable of the free Hamiltonian. 
So far, we have quantized the electromagnetic field with respect to the Minkowski time $t$. 
The free Hamiltonian for the field is therefore a generator of time-translation with respect to $t$, and we indicate this by writing 
\begin{align}
    \hat H\ts{EM,0}^t
    &=
        \int_{\R^3} \dd^3 k
        \sum_{\lambda=1}^2
        \kk \hat a_{\bk,\lambda}^\dag \hat a_{\bk,\lambda}\,. \label{eq:field free Hamiltonian}
\end{align}
In the following sections, we consider the interaction between the field and a spin particle with proper time $\tau$, and we will express the full Hamiltonian in terms of $\tau$. 
Hence, we will use the field's free Hamiltonian as a generator of time-translation with respect to the proper time $\tau$, $\hat H\ts{EM,0}^\tau$, given by 
\begin{align}
    \hat H\ts{EM,0}^\tau
    &=
        \dfrac{\dd t(\tau)}{ \dd \tau} 
        \hat H\ts{EM,0}^t\,.
\end{align}
This can be understood as follows \cite{EMM.Relativistic.quantum.optics, Tales2020GRQO}. 
In general, the Schr\"odinger equation with respect to $t$ is $\ii \frac{\dd}{\dd t} \ket{\psi(t)}= \hat H^t(t) \ket{\psi(t)}$. 
By changing the time variable to $\tau$ using $t=t(\tau)$, we obtain the above relation: 
\begin{align}
    \ii \dfrac{\dd }{\dd \tau} \ket{\psi(t(\tau))}
    &=
        \dfrac{ \dd t(\tau) }{\dd \tau} \hat H^t(t(\tau)) \ket{\psi(t(\tau))}
    \equiv
        \hat H^\tau(\tau) \ket{\psi(\tau)}\,. \notag 
\end{align}

\subsection{Spin-magnetic field system}
Consider a spin-$\frac{1}{2}$ particle interacting with a quantum electromagnetic field. 
In this paper, we consider three degrees of freedom: the spin, the COM, and the quantum field. 
The Hilbert space is given by $\hil\ts{s} \otimes \hil\ts{COM} \otimes \hil\ts{EM}$, with `s' standing for the spin system. 
However, since the COM degree of freedom is only relevant for spatial superposition and is not involved in the interaction, we focus only on $\hil\ts{s} \otimes \hil\ts{EM}$ for now.

We first assume that the spin particle is exposed to a \textit{classical} magnetic field $\bm{B}(\bx)$, which is directed in the $+z$-axis, i.e., 
\begin{align}
    \bm{B}(\bx)
    &=
        [ 0, 0, B_0(\bx) ]^\intercal\,.
\end{align}
Note that the magnetic field is not uniform in general, thereby it could depend on $\bx$. 
In the Schr\"odinger picture, the free Hamiltonian for the spin system $\hat H\ts{s,0}^\tau$ as a generator of the time-translation with respect to the spin's proper time $\tau$ is given by 
\begin{align}
    \hat H\ts{s,0}^\tau
    &=
        \hat{\bm{\mu}} \cdot \bm{B}(\bx)
    =
        \mu\ts{B} B_0(\bx) \hat \sigma_3 \in \mathcal{B}(\hil\ts{s})\,, \label{eq:spin free Hamiltonian}
\end{align}
where $\mu\ts{B}$ is the Bohr magneton and $\hat{\bm{\mu}} \equiv \mu\ts{B} \hat{\bm{\sigma}}$ is the magnetic moment operator defined with the vector of the Pauli operators $\hat{\bm{\sigma}}=[ \hat \sigma_1, \hat \sigma_2, \hat \sigma_3 ]^\intercal$. 
Here, $\mathcal{B}(\hil\ts{s})$ denotes the space of bounded operators on the spin's Hilbert space. 
It is worth noting that the Bohr magneton $\mu\ts{B}$ has the dimension of length in natural units. 
We define $\ket{\uparrow}$ and $\ket{\downarrow}$ as the eigenstates of $\hat H\ts{s,0}^\tau$ (or equivalently, eigenstates of $\hat \sigma_3$): 
\begin{align}
    \hat H\ts{s,0}^\tau \ket{\uparrow}
    &=
        \mu\ts{B} B_0(\bx) \ket{\uparrow}\,,
    \quad
    \hat H\ts{s,0}^\tau \ket{\downarrow}
    =
        -\mu\ts{B} B_0(\bx) \ket{\downarrow}\,.
\end{align}

Let us now consider the interaction between the spin system and a quantized magnetic field $\hat{\bm{B}}(\sx)$. 
In the Schr\"odinger picture, the total Hamiltonian $\hat H\ts{S,tot}$ is given by 
\begin{align}
    \hat H\ts{S,tot}^\tau
    &=
        \hat H\ts{s,0}^\tau \otimes \id\ts{EM} 
        + 
        \id\ts{s} \otimes \hat H\ts{EM,0}^\tau
        + \hat H\ts{S,int}^\tau (\tau)\,,
\end{align}
where 
\begin{align}
    \hat H\ts{S,int}^\tau(\tau)
    &=
        \hat{\bm{\mu}}(\tau) \cdot \hat{\bm{B}}(\sx)
    =
        \mu\ts{B} \chi(\tau) \hat \sigma_i \otimes \hat B^i(\sx)\,.
\end{align}
Here, $\tau$ is the proper time of the spin particle, $\chi(\tau)$ is the switching function that governs the time dependence of the interaction, and the Einstein summation convention is used for $i=1,2,3$. 
The switching function $\chi(\tau)$ is typically assumed to be a compactly supported smooth function, $\chi\in C_0^\infty (\R)$. 
In the literature, it is also common that $\chi(\tau)$ is chosen to be a smooth superpolynomial function such as a Gaussian function.

Let us write the sum of free Hamiltonians as $\hat H\ts{0}^\tau \equiv \hat H\ts{s,0}^\tau+ \hat H\ts{EM,0}^\tau$. 
The interaction Hamiltonian $\hat H\ts{I}^\tau$ in the interaction picture is then \cite{Ford.Macroscopic.1992, Doukas.spin.entanglement.2009}
\begin{align}
    \hat H\ts{I}^\tau
    &=
        e^{ \ii \hat H\ts{0}^\tau \tau } \hat H\ts{S,int}^\tau(\tau) e^{ -\ii \hat H\ts{0}^\tau \tau } \notag \\
    &=
        \mu\ts{B} \chi(\tau)
        \Bigkagikako{
            e^{ \ii \Omega(\bx) \tau } \hat \sigma_+
            \otimes 
            \hat B_-(\sx(\tau)) \notag \\
            &\quad
            +
            e^{ -\ii \Omega(\bx) \tau } \hat \sigma_-
            \otimes 
            \hat B_+(\sx(\tau))
            +
            \hat \sigma_3 
            \otimes 
            \hat B_3(\sx(\tau))
        }\,, \label{eq:interaction Hamiltonian in interaction pic}
\end{align}
where $\Omega(\bx)\coloneqq 2 \mu\ts{B} B_0(\bx)$ is the energy difference between the two spin states, $\hat \sigma_\pm$ are the raising and lowering operators $\hat \sigma_\pm \coloneqq (\hat \sigma_1 \pm \ii \hat \sigma_2)/2$ (or equivalently, $\hat \sigma_+ = \ket{\uparrow}\bra{\downarrow}$ and $\hat \sigma_- = \ket{\downarrow}\bra{\uparrow}$), and $\hat B_\pm(\sx)\coloneqq \hat B_1(\sx) \pm \ii \hat B_2(\sx)$. 
The first two terms involving $\hat \sigma_\pm$ in \eqref{eq:interaction Hamiltonian in interaction pic} correspond to interactions with a spin flip, while the third term does not. 
Hence, the first two terms are analogous to the Unruh-DeWitt particle detector model \cite{Unruh1979evaporation, DeWitt1979}, which is linearly coupled to a quantum scalar field.

The time-evolution operator $\hat U\ts{I}$ is given by 
\begin{align}
    \hat U\ts{I}
    &=
        \mathcal{T}_t 
        \exp 
        \kagikako{
            -\ii \int_\R \dd t\,
            \dfrac{\dd \tau}{\dd t} 
            \hat H\ts{I}^\tau(\tau(t))
        }\,,
\end{align}
where $\mathcal{T}_t$ is a time-ordering symbol with respect to time $t$. 
In what follows, we will assume that the energy difference $\Omega$ is constant during the interaction for simplicity: $\Omega(\bx)=\Omega$. 
This energy $\Omega$ can be thought of as a typical energy scale of our system. 
To use a perturbative analysis, we assume that the dimensionless quantity $\mu\ts{B} \Omega$ is much smaller than 1, i.e., $\mu\ts{B} \Omega \ll 1$.\footnote{
One needs to consider a dimensionless coupling constant to compare it to unity. 
} 
By performing the Dyson series expansion, we obtain 
\begin{align}
    \hat U\ts{I}
    &=
        \id 
        + \hat U\ts{I}^{(1)} 
        + \hat U\ts{I}^{(2)}
        + \cdots\,, \label{eq:perturbed unitary operator}
\end{align}
where 
\begin{subequations}
    \begin{align}
        \hat U\ts{I}^{(1)}
        &\coloneqq
            -\ii \int_\R \dd t\,\dfrac{\dd \tau}{\dd t} \hat H\ts{I}^\tau(\tau (t))\,, \label{eq:U1} \\
        \hat U\ts{I}^{(2)}
        &\coloneqq
            - \int_\R \dd t
            \int_\R \dd t'\,\Theta(t-t')
            \dfrac{\dd \tau}{\dd t}
            \dfrac{\dd \tau}{\dd t'} 
            \hat H\ts{I}^\tau(\tau(t)) \hat H\ts{I}^\tau(\tau(t'))\,, \label{eq:U2}
    \end{align}
\end{subequations}
with $\Theta(t)$ being the Heaviside step function. 
Here, $\hat U\ts{I}^{(j)}$ indicates that it contains the $j$th power of the Bohr magneton: $\mu\ts{B}^j$.

\section{Final density matrix}\label{sec:final density matrix}
Using perturbation theory, we derive the final density matrix for the spin particle. 
Here, we explicitly include the COM degree of freedom to consider the spin's spatial superposition state.

\subsection{Initial state}\label{subsec:initial state}

Let us assume that in the infinite past, the initial state of the total system is given by 
\begin{align}
    \ket{\psi\ts{s,0}} \otimes \ket{C} \otimes \ket{\text{EM}}\,, \notag 
\end{align}
where $\ket{C}$ is the COM state, $\ket{\psi\ts{s,0}}$ is the spin degree of freedom's initial pure state given by 
\begin{align}
    \ket{\psi\ts{s,0}}
    &=
        \alpha \ket{\uparrow} + e^{ \ii \vartheta } \sqrt{1-\alpha^2} \ket{\downarrow}\,,
\end{align}
where $\alpha \in [0,1]$ and $\vartheta \in [0, 2\pi)$ is the relative phase, and $\ket{\text{EM}}$ corresponds to the initial state of the field. 
We note that the field state $\ket{\text{EM}}$ is written in this way for the sake of convenience.

Before the spin particle interacts with the quantum field, we assume the classical magnetic field $B_0(\bx)$ causes the particle's trajectory to split based on its spin. 
To be precise, a Stern-Gerlach-type protocol entangles the COM state $\ket{C}$ with the spin degree of freedom: 
\begin{align}
    \ket{\psi\ts{s,0}}  \otimes \ket{C}
    \mapsto 
        \alpha \ket{\uparrow, C_\uparrow} + e^{\ii \vartheta} \sqrt{1-\alpha^2} \ket{\downarrow, C_\downarrow}  \,.
\end{align}
Here $\ket{C_\uparrow}$ and $\ket{C_\downarrow}$ represent the trajectories corresponding to the up-spin and down-spin, respectively. 
Furthermore, for simplicity, we assume that each wavepacket is very narrow so that $\{ \ket{C_\uparrow}, \ket{C_\downarrow} \}$ is orthonormal. 
As a result, the dimension of the joint Hilbert space of the spin and the COM after splitting is $2 \times 2$. 
In the basis $\{ \ket{\uparrow, C_\uparrow}, \ket{\uparrow, C_\downarrow}, \ket{\downarrow, C_\uparrow}, \ket{\downarrow, C_\downarrow} \}$, the initial joint density matrix for the spin particle is 
\begin{align}
    \rho\ts{s,0}
    &=
        \begin{bmatrix}
            \alpha^2 & 0 & 0 & \alpha \sqrt{1-\alpha^2} e^{-\ii \vartheta} \\
            0 & 0 & 0 & 0 \\
            0 & 0 & 0 & 0 \\
            \alpha \sqrt{1-\alpha^2} e^{\ii \vartheta} & 0 & 0 & 1-\alpha^2
        \end{bmatrix}\,,\label{eq:spin initial state}
\end{align}
and we will denote the total initial density matrix $\rho\ts{tot,0}$ by 
\begin{align}
    \rho\ts{tot,0}
    &=
        \rho\ts{s,0} \otimes \rho\ts{EM,0}\,,\label{eq:total initial state}
\end{align}
where $\rho\ts{EM,0}$ is the field's initial state.

\begin{figure*}[t]
\centering
\includegraphics[width=\linewidth]{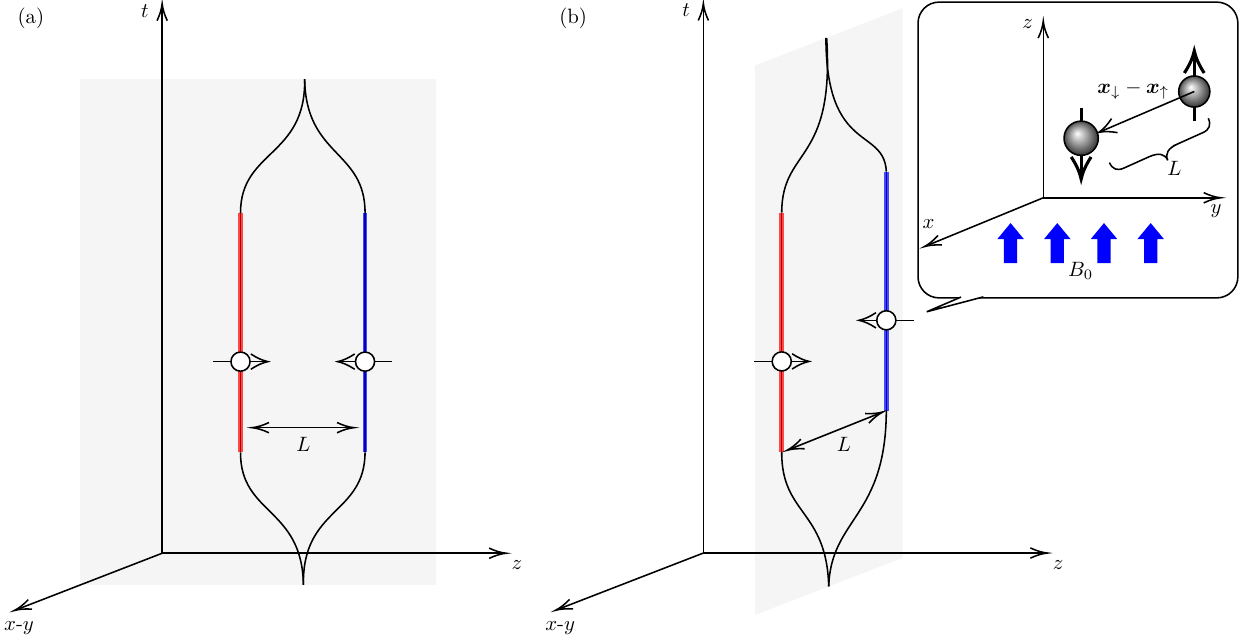}
\caption{Comparing the two cases. 
(a) The two superposed trajectories split in the $z$-direction and remain static relative to Minkowski time $t$. 
The red and blue lines indicate the duration of interaction. 
(b) The scenario where the trajectories split in the $(x,y)$-plane, i.e., their displacement vector is perpendicular to the $z$-axis. 
}
\label{fig:setup}
\end{figure*}

The spatial dependence of the classical magnetic field vector $B_0(\bx)$ allows for the possibility of manipulating the spin's trajectory based on its spin state. 
By ingeniously engineering this position dependence of $B_0(\bx)$, it becomes feasible to direct the spin's trajectory in such a way that the displacement vector, which characterizes the direction of split between trajectories, can either be aligned parallel to the $z$-axis or lie perpendicular to it, within the $(x,y)$-plane. 
This concept is illustrated in Fig.~\ref{fig:setup}.

Finally, let us specify the field's initial state $\rho\ts{EM,0}$. 
In this paper, we assume that the initial state of the field, $\rho\ts{EM,0}$, is a quasifree state. 
This is a special type of state where any connected $n$-point correlation function can be expressed as products of the Wightman function (i.e., two-point correlation function). 
To be precise, let us define the $n$-point correlation function as 
\begin{align}
    &\braket{\hat B_i(\sx_1) \hat B_j(\sx_2) \cdots \hat B_k(\sx_n)}_{\rho\ts{EM,0}} \notag \\
    &\coloneqq
        \Tr[ \hat B_i(\sx_1) \hat B_j(\sx_2) \cdots \hat B_k(\sx_n) \rho\ts{EM,0} ]\,,
\end{align}
where $\hat B_i(\sx)$ is the $i$th component of the quantum magnetic field $\hat{\bm{B}}(\sx)$ \eqref{eq:B mode expansion}. 
Then, $\rho\ts{EM,0}$ is said to be a quasifree state if any $n$-point correlation function can be written as products of the Wightman function $\braket{\hat B_i(\sx) \hat B_j(\sx')}_{\rho\ts{EM,0}}$. 
Examples of quasifree states include, but are not limited to, the Minkowski vacuum state, the squeezed state, and the thermal state known as the Kubo-Martin-Schwinger (KMS) state \cite{Kubo1957thermality, Martin-Schwinger1959thermality}.

\subsection{Final state}\label{subsec:final state}
Up to this point, the interaction between the spin and the field has not been turned on. 
We assume that, after we split the trajectories, the classical magnetic field is uniform and time-independent so that $\Omega$ can be treated as a constant throughout the interaction. 
We then smoothly turn on the interaction for duration $\sim \sigma$. 
After the interaction, we perform a reverse Stern-Gerlach-type procedure to converge the trajectories by admitting the spacetime dependence of $B_0(\bx)$.

Note that we will consider the interaction between the spin system and the quantum electromagnetic field while the spin particle remains at rest. 
This approach allows us to investigate the intrinsic effects of vacuum and thermal fluctuations on the spin superposition state. 
On the other hand, if we also consider the interaction during the splitting/recombination periods, there appears additional
decoherence on the spin state. 
Especially, the spin particle becomes entangled with the emitted photons due to acceleration radiation (see e.g., \cite{Sugiyama:2022ixw, Breuer.coherence.2001}).
However, such decoherence can be neglected by performing the splitting/recombination processes very slowly. 
Thus, to focus on the effects of vacuum fluctuation and thermal noise on the superposition state, we adopt the initial state Eq.~(\ref{eq:total initial state}) and simplify our calculations by considering the interaction only when the spin particle is at rest. 

Let us employ a perturbative method to obtain the final density matrix of the spin particle. 
Starting from the initial state \eqref{eq:total initial state} with \eqref{eq:spin initial state}, the entire system evolves under the unitary operator $\hat U\ts{I}$ given by \eqref{eq:perturbed unitary operator} with \eqref{eq:U1} and \eqref{eq:U2}. 
The final density matrix of the spin particle $\rho\ts{s}$ (i.e., the spin and COM degrees of freedom) after the interaction can be obtained by tracing out the field degree of freedom from the final total state as follows. 
\begin{align}
    \rho\ts{s}
    &=
        \Tr\ts{EM}
        [ \hat U\ts{I} \rho\ts{tot,0} \hat U\ts{I}^\dag ] \notag \\
    &=
        \rho\ts{s}^{(0,0)}
        + 
        \rho\ts{s}^{(1,1)}
        + 
        \rho\ts{s}^{(2,0)}
        + 
        \rho\ts{s}^{(0,2)}
        + \mathcal{O}(\mu\ts{B}^4)\,,
\end{align}
where $\Tr\ts{EM}$ is understood as a partial trace of the field degree of freedom, and 
\begin{align}
    \rho\ts{s}^{(i,j)}
    &\coloneqq
        \Tr\ts{EM}[ \hat U\ts{I}^{(i)} \rho\ts{tot,0} \hat U\ts{I}^{(j)\dagger} ]
\end{align}
with $\rho\ts{s}^{(0,0)}\equiv \rho\ts{s,0}$. 
Note that $\rho\ts{s}$ includes only even powers of $\mu\ts{B}$ because the odd-point correlation functions vanish for quasifree states.

Since we are considering the spatial superposition state, it is necessary to take into account the spacetime point at which the field observables $\hat B_i(\sx)$ interact. 
Recall that each $\hat U\ts{I}^{(i)}$ contains the interaction Hamiltonians $\hat H\ts{I}^\tau(\tau)$ \eqref{eq:interaction Hamiltonian in interaction pic}, each of which takes the form $\hat \sigma_i \otimes \hat B_i(\sx)$. 
We impose that the field observable is defined on $\sx_\uparrow$ (resp. $\sx_\downarrow$) when $\hat \sigma_i \otimes \hat B_i(\sx)$ acts on the states associated with $\ket{C_\uparrow}$ (resp. $\ket{C_\downarrow}$), that is, for any $s \in \{ \uparrow, \downarrow \}$, 
\begin{align}
    &[\hat \sigma_i \otimes \hat B_i(\sx)] 
    (\ket{s, C_\uparrow} \otimes \ket{\text{EM}})
    \mapsto 
    \hat \sigma_i \ket{s, C_\uparrow} \otimes \hat B_i(\sx_\uparrow) \ket{\text{EM}}\,, \notag \\
    &[\hat \sigma_i \otimes \hat B_i(\sx)] 
    (\ket{s, C_\downarrow} \otimes \ket{\text{EM}})
    \mapsto 
    \hat \sigma_i \ket{s, C_\downarrow} \otimes \hat B_i(\sx_\downarrow) \ket{\text{EM}}\,. \notag
\end{align}

The resulting density matrix $\rho\ts{s}$ reads
\begin{align}
    \rho\ts{s}
    &=
        \begin{bmatrix}
            \rho_{11} & 0 & 0 & \rho_{14} \\
            0 & \rho_{22} & \rho_{23} & 0 \\
            0 & \rho_{23}^* & \rho_{33} & 0 \\
            \rho_{14}^* & 0 & 0 & \rho_{44} 
        \end{bmatrix} 
        + \mathcal{O}(\mu\ts{B}^4)\,, \label{eq:final density 4x4}
\end{align}
where 
\begin{subequations}
    \begin{align}
        \rho_{11}
        &= 
            \rho_{11}^{(0)} 
            [
                1 - P_{\uparrow \to \downarrow}^{C_\uparrow}(\Omega)
            ]\,,  \\
        \rho_{22}
        &= 
            \rho_{44}^{(0)} P_{\downarrow \to \uparrow}^{C_\downarrow}(\Omega)\,,\\
        \rho_{33}
        &= 
            \rho_{11}^{(0)}
            P_{\uparrow \to \downarrow}^{C_\uparrow}(\Omega)\,, \\
        \rho_{44} 
        &= 
            \rho_{44}^{(0)}
            [
                1- P_{\downarrow \to \uparrow}^{C_\downarrow}(\Omega)
            ] \,, \\
        \rho_{14}
        &= 
            \rho_{14}^{(0)} (1 - \mathcal{D}\ts{nl} - \mathcal{A}\ts{loc} - \mathcal{D}\ts{loc} )\,, \\
        \rho_{23}
        &=
            \rho_{14}^{(0)*}
            \mathcal{M}\ts{nl}\,, 
    \end{align}\label{eq:elements of final state 4x4}
\end{subequations}
where $\rho_{11}^{(0)}=\alpha^2, \rho_{44}^{(0)}=1-\alpha^2$, and $\rho_{14}^{(0)}=\alpha \sqrt{1-\alpha^2} e^{-\ii \vartheta}$, and 
\begin{subequations}
    \begin{align}
    &P_{\downarrow \to \uparrow}^{C_\downarrow}(\Omega)
    \coloneqq
        \int_\R \dd t 
            \int_\R \dd t'\,
            \braket{\mathscr{B}_+(\sx_\downarrow) \mathscr{B}_-(\sx'_\downarrow)}_{\rho\ts{EM,0}}\,, \\
    &P_{\uparrow \to \downarrow}^{C_\uparrow}(\Omega)
    \coloneqq
        \int_\R \dd t 
            \int_\R \dd t'\,
            \braket{\mathscr{B}_-(\sx_\uparrow) \mathscr{B}_+(\sx'_\uparrow)}_{\rho\ts{EM,0}}\,, \\
    &\mathcal{M}\ts{nl}
    \coloneqq
        \int_\R \dd t 
            \int_\R \dd t'\,
            \braket{\mathscr{B}_-(\sx_\uparrow) \mathscr{B}_-(\sx'_\downarrow)}_{\rho\ts{EM,0}}\,,\\
    &\mathcal{D}\ts{nl}
    \coloneqq
        \int_\R \dd t 
            \int_\R \dd t'\,
            \braket{\mathscr{B}_3(\sx_\downarrow) \mathscr{B}_3(\sx'_\uparrow)}_{\rho\ts{EM,0}}\,, \\
    &\mathcal{A}\ts{loc}
    \coloneqq
            \int_\R \dd t 
            \int_\R \dd t'\,\Theta(t-t')
            \braket{\mathscr{B}_-(\sx_\uparrow) \mathscr{B}_+(\sx'_\uparrow)}_{\rho\ts{EM,0}} \notag \\
            &\quad
            +\int_\R \dd t 
            \int_\R \dd t'\,\Theta(t'-t)
            \braket{\mathscr{B}_+(\sx_\downarrow) \mathscr{B}_-(\sx'_\downarrow)}_{\rho\ts{EM,0}}\,, \\
    &\mathcal{D}\ts{loc} 
    \coloneqq
            \int_\R \dd t 
            \int_\R \dd t'\,\Theta(t-t')
            \braket{\mathscr{B}_3(\sx_\uparrow) \mathscr{B}_3(\sx'_\uparrow)}_{\rho\ts{EM,0}}  \notag \\
            &\quad
            +
            \int_\R \dd t 
            \int_\R \dd t'\,\Theta(t'-t)
            \braket{\mathscr{B}_3(\sx_\downarrow) \mathscr{B}_3(\sx'_\downarrow)}_{\rho\ts{EM,0}}\,,
\end{align}
\end{subequations}
and 
\begin{subequations}
    \begin{align}
        \hat{\mathscr{B}}_\mp (\sx(t))
        &\coloneqq
            \mu\ts{B} \dfrac{\dd \tau}{\dd t} \chi(\tau(t)) e^{ \pm \ii \Omega \tau(t) } \hat B_\mp(\sx(\tau(t)))\,, \\
        \hat{\mathscr{B}}_3 (\sx(t))
        &\coloneqq
            \mu\ts{B} \dfrac{\dd \tau}{\dd t} \chi(\tau(t)) \hat B_3(\sx(\tau(t)))\,.
    \end{align}
\end{subequations}
We note that the terms with $\braket{\hat B_3 \hat B_\pm}_{\rho\ts{EM,0}}$ vanish.

Here, $P^{C_\downarrow}_{\downarrow \to \uparrow}(\Omega)$ [resp. $P^{C_\uparrow}_{\uparrow \to \downarrow}(\Omega)$] is the transition probability of the spin degree of freedom from $\ket{\downarrow}$ to $\ket{\uparrow}$ (resp. from $\ket{\uparrow}$ to $\ket{\downarrow}$). 
The superscript $C_s$ with $s\in \{ \uparrow, \downarrow \}$ indicates the path along which it is evaluated. 
In other words, the transition probabilities are obtained by evaluating the two-point correlation functions along each corresponding trajectory: $\braket{ \hat B_+(\sx_\downarrow) \hat B_-(\sx'_\downarrow) }_{\rho\ts{EM,0}}$ or $\braket{ \hat B_-(\sx_\uparrow) \hat B_+(\sx'_\uparrow) }_{\rho\ts{EM,0}}$. 
Henceforth, we will refer to such quantities that depend on a single trajectory as ``local contributions.'' 
Other local contributions are $\mathcal{A}\ts{loc}$ and $\mathcal{D}\ts{loc}$. 
We will see in the next section that the real parts of these local contributions, $\text{Re}[\mathcal{A}\ts{loc}]$ and $\text{Re}[\mathcal{D}\ts{loc}]$, correspond to decoherence terms. 
They can be written as 
\begin{align}
    \text{Re}[ \mathcal{A}\ts{loc} ]
    &=
        \dfrac{1}{2} 
        [
            P_{\uparrow \to \downarrow}^{C_\uparrow}(\Omega)
            +
            P_{\downarrow \to \uparrow}^{C_\downarrow}(\Omega)
        ]\,, \label{eq:real amplitude damp local} \\
    \text{Re}[ \mathcal{D}\ts{loc} ]
    &=
        \dfrac{1}{2}
        \bigg(
        \int_\R \dd t 
            \int_\R \dd t'\,
            \braket{\mathscr{B}_3(\sx_\uparrow) \mathscr{B}_3(\sx'_\uparrow)} \notag \\
        &\quad
        +
        \int_\R \dd t 
            \int_\R \dd t'\,
            \braket{\mathscr{B}_3(\sx_\downarrow) \mathscr{B}_3(\sx'_\downarrow)}
        \bigg) \,. \label{eq:real local deco phase damp}
\end{align}
We will call them the ``local decoherence terms.''

On the other hand, $\mathcal{M}\ts{nl}$ and $\mathcal{D}\ts{nl}$ are ``nonlocal'' since they depend on the two-point correlation functions across two different trajectories: $\braket{\hat B_- (\sx_\uparrow) \hat B_-(\sx_\downarrow')}_{\rho\ts{EM,0}}$ and $\braket{\hat B_3 (\sx_\downarrow) \hat B_3(\sx_\uparrow')}_{\rho\ts{EM,0}}$. 
In particular, $\mathcal D\ts{nl}$ only depends on the anticommutator (i.e., the Hadamard distribution) $\braket{\{ \hat B_3 (\sx_\downarrow), \hat B_3(\sx_\uparrow')\}}_{\rho\ts{EM,0}}$. 

Let us introduce the types of decoherence: \textit{amplitude damping} and \textit{phase damping} (dephasing). 
Amplitude damping is a process corresponding to the spontaneous emission. 
In our case, this is governed by the first two terms of the form $e^{\ii \Omega \tau} \hat \sigma_+ \hat B_- + e^{-\ii \Omega \tau} \hat \sigma_- \hat B_+$ in the interaction Hamiltonian \eqref{eq:interaction Hamiltonian in interaction pic}. 
The spin particle exchanges energy with the field and a spin flip occurs. 
In this way, the spin and the emitted photon become entangled, leading to the spin's decoherence. 
Amplitude damping, in general, alters both diagonal and off-diagonal elements in a density matrix, which is also the case for our density matrix \eqref{eq:final density 4x4}. 
The elements of the density matrix with $\hat B_\pm(\sx)$, such as the transition probabilities, the nonlocal term $\mathcal{M}\ts{nl}$, and the local decoherence term $\mathcal{A}\ts{loc}$ are the consequence of amplitude damping.

Phase damping, on the other hand, is governed by the term $\hat \sigma_3 \hat B_3$ in the interaction Hamiltonian \eqref{eq:interaction Hamiltonian in interaction pic}. 
Since $\hat \sigma_3$ does not flip the states $\ket{\uparrow}$ and $\ket{\downarrow}$, phase damping corresponds to the loss of the relative phase \cite{nielsen2000quantum} in the spin particle without exchanging energy with the field. 
Thus, a density matrix loses the off-diagonal elements only. 
In our final density matrix \eqref{eq:final density 4x4}, the elements $\mathcal{D}\ts{nl}$ and $\mathcal{D}\ts{loc}$ are the results of phase damping.

In the next section, we compute a decoherence measure. 
As we will see, the decoherence measure can be written with $\text{Re}[ \mathcal{A}\ts{loc} ]$, $\text{Re}[ \mathcal{D}\ts{loc} ]$, $\text{Re}[ \mathcal{D}\ts{nl} ]$, and $|\mathcal{M}\ts{nl}|$. 
Thus, let us first evaluate each of these by specifying the field's initial state $\rho\ts{EM,0}$ and the spin's trajectory $\sx(\tau)$.

\subsubsection{The Minkowski vacuum case}
Here, we assume that the field is initially in the Minkowski vacuum $\ket{0}$, and that the spin particle is at rest with respect to the Minkowski time $t$ so that the trajectory can be written as $\sx_s(\tau)=(\tau, \bx_s)$, where $\bx_s$ with $s\in \{ \uparrow, \downarrow \}$ is a fixed spatial vector describing the spatial position of the spin with the COM state $\ket{C_s}$, and $\tau$ is now understood as $\tau=t$.

We first evaluate the excitation transition probability $P^{C_\downarrow}_{\downarrow \to \uparrow}(\Omega)$. 
The same transition probability was examined in \cite{Ford.Macroscopic.1992}, but they encountered an ultraviolet (UV) divergence. 
We will comment on their results below.

Inserting the mode expansion given in \eqref{eq:B mode expansion}, one finds 
\begin{align}
    &\braket{\hat B_+(\sx(\tau)) \hat B_-(\sx(\tau')) }_{\rho\ts{EM,0}}
    =
        \braket{0|\hat B_+(\sx(\tau)) \hat B_-(\sx(\tau')) |0} \notag \\
    &=
        \int_{\R^3} \dd^3 k \sum_\lambda 
        B_{(\bk, \lambda)}^{(+)}(\sx(\tau)) B_{(\bk, \lambda)}^{(+)*}(\sx(\tau'))\,, \notag 
\end{align}
where $B_{(\bk, \lambda)}^{(+)}(\sx)\coloneqq B_{(\bk, \lambda)}^{(1)} (\sx)
                + \ii
                B_{(\bk, \lambda)}^{(2)} (\sx)$, 
and so the transition probability reduces to 
\begin{align}
    P^{C_\downarrow}_{\downarrow \to \uparrow}(\Omega)
    &=
        \mu\ts{B}^2
        \int_{\R^3} \dfrac{ \dd^3 k }{ (2\pi)^3 2\kk }
        (\kk^2 + k_3^2)
        |\tilde{\chi}(\kk + \Omega)|^2 \notag \\
    &=
        \dfrac{ \mu\ts{B}^2 }{ 3\pi^2 }
        \int_0^\infty \dd \kk\,\kk^3  |\tilde{\chi}(\kk + \Omega)|^2\,, \label{eq:transition prob final}
\end{align}
where we employed the Fourier transformation
\begin{align}
    \tilde{\chi}(\omega)
    &\coloneqq 
        \int_\R \dd \tau\,
        \chi(\tau) e^{ \ii \omega \tau }\,, \label{eq:Fourier trans}
\end{align}
and used the identity
\begin{align}
    \sum_\lambda (\bk \times \bm{e}_\lambda (\bk))_+ (\bk \times \bm{e}_\lambda^* (\bk))_-
    =
        \kk^2 + k_3^2\,,
\end{align}
where $(\bk \times \bm{e}_{(\bk, \lambda)})_\pm \equiv (\bk \times \bm{e}_{(\bk, \lambda)})_1 \pm \ii (\bk \times \bm{e}_{(\bk, \lambda)})_2$. 
This identity can be directly proven by using the projector \eqref{eq:projector}. 
Note that we have not specified the switching function $\chi(\tau)$ yet in \eqref{eq:transition prob final}.

We first comment on the results obtained by \cite{Ford.Macroscopic.1992}. 
The authors in \cite{Ford.Macroscopic.1992} considered a special case of our result \eqref{eq:transition prob final} when a rectangular switching function is chosen. 
In general, the Fourier transform of the rectangular function with width $2\sigma$ is known to be $2\sin(\omega \sigma)/\omega$. 
The integral in \eqref{eq:transition prob final} takes the form 
\begin{align}
    \int_0^\infty \dd \kk\,
    \dfrac{ \kk^3 \sin^2 [(\kk + \Omega) \sigma] }{ (\kk + \Omega)^2 }\,,
\end{align}
which is UV divergent, even though the switching function is compactly supported. 
This divergence is attributed to the poor choice of a switching function, which does not converge at large $\kk$. 
One can instead choose a more appropriate switching function, such as a Gaussian function, which ensures convergence: 
\begin{align}
    \chi(\tau)
    &=
        e^{ -\tau^2 / 2\sigma^2 }\,, \label{eq:Gaussian switching}
\end{align}
where $\sigma$ is the typical time scale. 
The Fourier transform of such a Gaussian function reads 
\begin{align}
    \tilde \chi(\omega)
    &=
        \sqrt{2\pi} \sigma e^{ -\omega^2 \sigma^2/2 }\,,
\end{align}
which is also a Gaussian, and the integrand in the transition probability \eqref{eq:transition prob final} converges at large $\kk$. 
The resulting transition probability reads 
\begin{align}
    P^{C_\downarrow}_{\downarrow \to \uparrow}(\Omega)
    &=
        \dfrac{\mu\ts{B}^2 }{ 6\pi \sigma^2 }
        \Bigkagikako{
            2 e^{ -\Omega^2 \sigma^2 } ( 1 + \Omega^2 \sigma^2 ) \notag \\
            &\quad
            - 
            \sqrt{\pi}
            \Omega \sigma 
            (3 + 2 \Omega^2 \sigma^2)
            \erfc(\Omega \sigma)
        }\,, \label{eq:transition prob Gaussian}
\end{align}
where $\erfc(x)$ is the complementary error function. 
The deexcitation transition probability $P^{C_\uparrow}_{\uparrow \to \downarrow}(\Omega)$ is obtained as 
\begin{align}
    P^{C_\uparrow}_{\uparrow \to \downarrow}(\Omega)
    &=
        \dfrac{\mu\ts{B}^2 }{ 6\pi \sigma^2 }
        \Bigkagikako{
            2 e^{ -\Omega^2 \sigma^2 } ( 1 + \Omega^2 \sigma^2 ) \notag \\
            &\quad
            + 
            \sqrt{\pi}
            \Omega \sigma 
            (3 + 2 \Omega^2 \sigma^2)
            \erfc(-\Omega \sigma)
        }\,, \label{eq:deexcitation transition prob Gaussian}
\end{align}
which is just $P^{C_\downarrow}_{\downarrow \to \uparrow}(-\Omega)$. 
Adding these transition probabilities yields the (vacuum) local amplitude damping term $\text{Re}[ \mathcal{A}\ts{loc} ]$ in \eqref{eq:real amplitude damp local}.

In the same manner, by using the identity 
\begin{align}
    \sum_{\lambda}
            (\bk \times \bm{e}_{(\bk, \lambda)})_3 
            (\bk \times \bm{e}_{(\bk, \lambda)}^*)_3 
        =
            \kk^2 - k_3^2\,,
\end{align}
the local phase damping term $\text{Re}[\mathcal{D}\ts{loc}]$ reads 
\begin{align}
    \text{Re}[\mathcal{D}\ts{loc}]
    &=
        \dfrac{ \mu\ts{B}^2 }{ 6\pi \sigma^2 }\,.\label{eq:vacuum local deco phase damp}
\end{align}

The nonlocal terms $\text{Re}[\mathcal{D}\ts{nl}]$ and $|\mathcal{M}\ts{nl}|$ depend on the configuration of the spin's spatial splitting. 
Let $\bx_\downarrow - \bx_\uparrow$ be a displacement vector pointing from the point on $\sx_\uparrow(\tau)$ to the point on $\sx_\downarrow(\tau)$ at a fixed time $\tau$. 
Since we are assuming static superposed trajectories, the displacement vector is time-independent. 
The length of the displacement vector, $L\equiv |\bx_\downarrow - \bx_\uparrow|$, represents the spatial distance between the two trajectories. 
Then, as we will show below, one can consider two scenarios: (i) when the displacement vector is parallel to the $z$-axis [see Fig.~\ref{fig:setup}(a)]; and (ii) when the displacement vector is perpendicular to the $z$-axis [see Fig.~\ref{fig:setup}(b)]. 
We provide a detailed calculation in Appendix~\ref{app:elements in coherence}.

For the nonlocal phase damping term $\text{Re}[\mathcal{D}\ts{nl}]$, a straightforward calculation leads to: \\
(i) When the spin is split along the $z$-axis: $\bx_\downarrow - \bx_\uparrow=[0,0,L]^\intercal$,
\begin{align}
    \text{Re}[\mathcal{D}\ts{nl}]
    &=
        \dfrac{\mu\ts{B}^2 \sigma }{ 2\pi L^3 } 
        \kagikako{
            -\dfrac{L}{\sigma}
            +
            \kako{
                \dfrac{L^2}{\sigma^2}+2
            }
            D^+
            \kako{
                \dfrac{L}{2\sigma}
            }
        } \label{eq:vacuum nonlocal dephasing z-split} \,, 
\end{align}
where $D^+(x)\coloneqq \sqrt{\pi} e^{-x^2} \erfi(x)/2$ is the Dawson function with $\erfi(x)$ being the imaginary error function. 
In particular, if the spatial splitting is small, it reduces to 
\begin{align}
    \text{Re}[\mathcal{D}\ts{nl}]
    &\simeq 
        \dfrac{\mu\ts{B}^2 }{ 6\pi \sigma^2 } 
        \quad ({\rm for } ~L\ll \sigma) \,. \label{eq:vac nonlocal phase damp z}
\end{align}
(ii) When the spin is split along the $x$-axis: $\bx_\downarrow - \bx_\uparrow=[L,0,0]^\intercal$,
\begin{align}
    \text{Re}[\mathcal{D}\ts{nl}]
    &=
        \dfrac{\mu\ts{B}^2 \sigma }{ 8\pi L^3 } 
        \kagikako{
            \dfrac{L^3}{\sigma^3}
            +2 \dfrac{L}{\sigma}
            -
            \kako{
                \dfrac{L^4}{\sigma^4}+4
            }
            D^+
            \kako{
            \dfrac{L}{2\sigma}
            }
        }\,, \label{eq:vacuum nonlocal dephasing x-split}         
\end{align}
which also reduces to 
\begin{align}
    \text{Re}[\mathcal{D}\ts{nl}]
    &\simeq 
        \dfrac{\mu\ts{B}^2 }{ 6\pi \sigma^2 } 
        \quad ({\rm for } ~L\ll \sigma) \,.\label{eq:vac nonlocal phase damp x}
\end{align}
While $\text{Re}[\mathcal{D}\ts{nl}]$ is nonnegative for case (i), it can be negative for case (ii) in some regions in a parameter space. 
Nevertheless, the amount of decoherence coincides with the local phase damping \eqref{eq:vacuum local deco phase damp} when the spatial displacement is small.

For the other nonlocal term $|\mathcal{M}\ts{nl}|$, we find: \\
(i) When the spin is split along the $z$-axis: $\bx_\downarrow - \bx_\uparrow=[0,0,L]^\intercal$,
\begin{align}
    |\mathcal{M}\ts{nl}|
    &=
        0\,;\label{eq:Mnl z-split}
\end{align}
(ii) When the spin is split along the $x$-axis: $\bx_\downarrow - \bx_\uparrow=[L,0,0]^\intercal$,
\begin{align}
    &|\mathcal{M}\ts{nl}|
    = \dfrac{ -3\mu\ts{B}^2 \sigma e^{-\Omega^2 \sigma^2} }{ 4\pi L^3 } \notag \\
        &\times 
        \kagikako{
                \dfrac{L}{\sigma} 
                \kako{
                    \dfrac{L^2}{6\sigma^2} + 1
                }
                -2
                \kako{
                    \dfrac{L^4}{12\sigma^4}
                    + \dfrac{L^2}{ 3\sigma^2 }
                    + 1
                }
                D^+
                \kako{
                    \dfrac{L}{2\sigma}
                }
            }\,,\label{eq:Mnl x-split}
\end{align}
which reduces to 
\begin{align}
    |\mathcal{M}\ts{nl}|
    \simeq
        \dfrac{ \mu\ts{B}^2 L^2 e^{-\Omega^2 \sigma^2} }{ 30\pi \sigma^4 }  
        \quad ({\rm for } ~L\ll \sigma) \,. \label{eq:Mnl x-split approx}
\end{align}

Let us comment on the rotation symmetry of the system. 
Although we considered the perpendicular scenario (ii) with a specific displacement vector $\bx_\downarrow - \bx_\uparrow=[L, 0, 0]^\intercal$, it is worth pointing out that the whole composite system is invariant under rotation about the $z$-axis. 
Therefore, the result does not change for $\bx_\downarrow - \bx_\uparrow=[L \cos \theta_3, L \sin \theta_3, 0]^\intercal$ as long as the quantum field is also rotated. 
To see this, it is straightforward to show that $[\hat H\ts{S,tot}, \hat U\ts{s}(\theta_3)\otimes \hat U\ts{EM}(\theta_3)]$=0, where $\hat H\ts{S,tot}$ is the full Hamiltonian in the Schr\"odinger picture, and $\hat U\ts{s}(\theta_3)(=e^{ 
-\ii \theta_3 \hat \sigma_3/2 })$ and $\hat U\ts{EM}(\theta_3)$ are the unitary representations of rotation about the $z$-axis for the spin particle and the quantum field, respectively. 
This commutation relation can be computed by realizing that $\hat U\ts{s}^\dag (\theta_3) \hat{\bm \sigma} \hat U\ts{s}(\theta_3)= D(\theta_3) \hat{\bm \sigma}$ as well as $\hat U\ts{EM}^\dag(\theta_3) \hat{\bm B}(\sx) \hat U\ts{EM}(\theta_3)=D(\theta_3) \hat{\bm B}(R^{-1}(\theta_3) \sx)$, where 
\begin{align}
    D(\theta_3)
    &=
        \begin{bmatrix}
            \cos \theta_3 & -\sin \theta_3 & 0 \\
            \sin \theta_3 & \cos \theta_3 & 0 \\
            0 & 0 & 1 \\
        \end{bmatrix}
\end{align}
is the rotation matrix and $R(\theta_3)$ is the rotation matrix for a spacetime point. 
These relations allow us to show that the $\hat{\bm \sigma} \cdot \hat{\bm B}(\sx)$ term in the interaction Hamiltonian is invariant under the joint unitary transformation $\hat U\ts{s}(\theta_3) \otimes \hat U\ts{EM}(\theta_3)$. 
On the other hand, if one rotates only the spin particle, i.e., $\hat U\ts{s}(\theta_3) \otimes \id\ts{EM}$, the entire system is not invariant. 
This leads to $\theta_3$-dependent results for $\bx_\downarrow - \bx_\uparrow=[L \cos \theta_3, L \sin \theta_3, 0]^\intercal$.

\subsubsection{Thermal state}
We are also interested in the case where the quantum field is in a thermal state. 
In quantum field theory, the thermal state is identified as the Kubo-Martin-Schwinger (KMS) state \cite{Kubo1957thermality, Martin-Schwinger1959thermality}. 
Technically, the Gibbs thermal state at inverse temperature $\beta$, 
\begin{align}
    \rho_\beta
    &=
        \dfrac{1}{Z} e^{ -\beta \hat H\ts{EM,0}^t }\,,\label{eq:Gibbs state}
\end{align}
where $Z\coloneqq \Tr[ e^{ -\beta \hat H\ts{EM,0}^t } ]$ is the partition function, is not well-defined for quantum fields in free space. 
This stems from the fact that the field's Hilbert space has uncountably many dimensions. 
Nevertheless, as discussed in Ref.~\cite{simidzija2018harvesting}, we can consider a large finite-sized box to make the Hilbert space countable and then take the infinite-size limit. 
This allows us to practically compute the correlation functions using the Gibbs state \eqref{eq:Gibbs state}, and the final result will agree with the rigorous definition of the KMS thermal state.

It is straightforward to see that one-point correlation functions vanish for the thermal state, and using Eqs.~\eqref{eq:Gibbs state} and \eqref{eq:field free Hamiltonian}, we obtain \cite{simidzija2018harvesting}
\begin{subequations}
\begin{align}
    &\Tr[ \rho_\beta \hat a_{\bk, \lambda} \hat a_{\bk', \lambda'}^\dag ]
    =
        \dfrac{ e^{ \beta \kk } }{ e^{\beta \kk} -1 } \delta^{(3)}(\bk - \bk') \delta_{\lambda, \lambda'}\,,\\
    &\Tr[ \rho_\beta \hat a_{\bk, \lambda}^\dag \hat a_{\bk', \lambda'} ]
    =
        \dfrac{ 1 }{ e^{\beta \kk} -1 } \delta^{(3)}(\bk - \bk') \delta_{\lambda, \lambda'}\,, \\
    &\Tr[ \rho_\beta \hat a_{\bk, \lambda} \hat a_{\bk', \lambda'} ]
    =
    \Tr[ \rho_\beta \hat a_{\bk, \lambda}^\dag \hat a_{\bk', \lambda'}^\dag ]
    =0\,.
\end{align}
\end{subequations}
From these properties, any thermal two-point correlation functions $\braket{\hat B_i \hat B_j}\ts{th}$ can be written as 
\begin{align}
    \braket{\hat B_i \hat B_j}\ts{th}
    &\equiv 
        \braket{\hat B_i \hat B_j}_{\rho_\beta}
    \coloneqq
        \Tr[ \rho_\beta \hat B_i \hat B_j ]
    \notag \\
    &=
        \braket{\hat B_i \hat B_j}\ts{vac}
        + 
        \braket{\hat B_i \hat B_j}_\beta\,, \label{eq:thermal total Wightman}
\end{align}
where $\braket{\hat B_i \hat B_j}\ts{vac}\coloneqq \braket{0|\hat B_i \hat B_j|0}$ is the vacuum two-point correlation function and 
\begin{align}
    \braket{\hat B_i \hat B_j}_\beta
    &\coloneqq 
        \int_{\R^3} \dd^3 k
        \sum_{\lambda}
        \dfrac{ B^{(i)}_{(\bk, \lambda)} B^{(j)*}_{(\bk, \lambda)} + \text{c.c.} }{ e^{\beta \kk } -1 }
\end{align}
is the contribution coming from the thermal state. 
Therefore, any quantities that we have obtained so far, such as transition probabilities, can be decomposed into the vacuum and thermal contributions: 
\begin{align}
    P_{\downarrow \to \uparrow}^{(\text{th})}=P_{\downarrow \to \uparrow}^{(\text{vac})} + P_{\downarrow \to \uparrow}^{(\beta)}\,.
\end{align}
Here, $P_{\downarrow \to \uparrow}^{(\text{vac})}$ and $P_{\downarrow \to \uparrow}^{(\beta)}$ come from $\braket{\hat B_+ \hat B_-}\ts{vac}$ and $\braket{\hat B_+ \hat B_-}_\beta$ in \eqref{eq:thermal total Wightman}, respectively. 
The excitation probability in the vacuum, $P_{\downarrow \to \uparrow}^{(\text{vac})}$, is given by \eqref{eq:transition prob Gaussian} and 
\begin{align}
    P_{\downarrow \to \uparrow}^{(\beta)}
    &=
        P_{\uparrow \to \downarrow}^{(\beta)}
    =
        \dfrac{\mu\ts{B}^2}{2(2\pi)^3} 
        \int_{\R^3} \dfrac{ \dd^3 k }{ \kk( e^{\beta \kk } -1 ) }
        (\kk^2 + k_3^2) \notag \\
        &\quad \times 
        \kako{
            |\tilde \chi(\Omega + \kk)|^2
            +
            |\tilde \chi(\Omega - \kk)|^2
        } \label{eq:transition prob thermal Fourier} \\
    &=
        \dfrac{\mu\ts{B}^2}{3\pi^2}
        \int_0^\infty \dd \kk 
        \dfrac{\kk^3}{ e^{\beta \kk} -1 } \notag  \\
        &\quad \times 
        \kako{
            |\tilde \chi(\Omega + \kk)|^2
            +
            |\tilde \chi(\Omega - \kk)|^2
        }\,. \notag
\end{align}
Employing the Gaussian switching function \eqref{eq:Gaussian switching}, 
\begin{align}
    &P_{\downarrow \to \uparrow}^{(\beta)}
    =
        P_{\uparrow \to \downarrow}^{(\beta)} \notag \\
    &=
        \dfrac{4 \mu\ts{B}^2 \sigma^2 e^{ -\Omega^2 \sigma^2 }}{3\pi}
        \int_0^\infty \dd \kk 
        \dfrac{\kk^3 e^{ -\kk^2 \sigma^2 }}{ e^{\beta \kk} -1 } 
        \cosh (2\Omega \kk \sigma^2)
        \,. 
\end{align}
Therefore, the local decoherence due to amplitude damping $\text{Re}[ \mathcal{A}\ts{loc}^{(\text{th})} ]$ in \eqref{eq:real amplitude damp local} becomes 
\begin{align}
    \text{Re}[ \mathcal{A}\ts{loc}^{(\text{th})} ]
    &=
        \text{Re}[ \mathcal{A}\ts{loc}^{(\text{vac})} ] + \text{Re}[ \mathcal{A}\ts{loc}^{(\beta)} ] \,,
\end{align}
where 
\begin{align}
    &\text{Re}[ \mathcal{A}\ts{loc}^{(\beta)} ]
    =
        (P_{\downarrow \to \uparrow}^{(\beta)} + P_{\uparrow \to \downarrow}^{(\beta)})/2 \notag  \\
    &=
        \dfrac{4 \mu\ts{B}^2 \sigma^2 e^{ -\Omega^2 \sigma^2 }}{3\pi}
        \int_0^\infty \dd \kk 
        \dfrac{\kk^3 e^{ -\kk^2 \sigma^2 }}{ e^{\beta \kk} -1 } 
        \cosh (2\Omega \kk \sigma^2)
        \,. \label{eq:beta amp damp local}
\end{align}
Similarly, the local decoherence due to phase damping $\text{Re}[ \mathcal{D}\ts{loc}^{(\text{th}) } ]$ in \eqref{eq:real local deco phase damp} reads
\begin{align}
    \text{Re}[ \mathcal{D}\ts{loc}^{(\text{th}) } ]
    &=
        \text{Re}[ \mathcal{D}\ts{loc}^{(\text{vac}) } ] + \text{Re}[ \mathcal{D}\ts{loc}^{(\beta) } ]\,,
\end{align}
where $\text{Re}[ \mathcal{D}\ts{loc}^{(\text{vac}) } ]$ is given in \eqref{eq:vacuum local deco phase damp} and 
\begin{align}
    \text{Re}[ \mathcal{D}\ts{loc}^{(\beta) } ]
    &=
        \dfrac{\mu\ts{B}^2}{(2\pi)^3} 
        \int_{\R^3} \dd^3 k
        \dfrac{ \kk^2 - k_3^2 }{ \kk (e^{\beta \kk} -1) } |\tilde \chi(\kk)|^2 \label{eq:local phase damp general Fourier} \\
    &=
        \dfrac{ 2\mu\ts{B}^2 \sigma^2 }{3\pi}
        \int_0^\infty \dd \kk 
        \dfrac{\kk^3 e^{ -\kk^2 \sigma^2 }}{ e^{\beta \kk} -1 }\,.
\end{align}

The nonlocal contributions, 
\begin{align}
    \text{Re}[\mathcal{D}\ts{nl}^{(\text{th})}]
    &=
        \text{Re}[\mathcal{D}\ts{nl}^{(\text{vac})}]
        +
        \text{Re}[\mathcal{D}\ts{nl}^{(\beta)}]\,, \\
    \mathcal{M}\ts{nl}^{(\text{th})}
    &=
        \mathcal{M}\ts{nl}^{(\text{vac})}
        +
        \mathcal{M}\ts{nl}^{(\beta)}\,,
\end{align}
where 
\begin{align}
    \text{Re}[\mathcal{D}\ts{nl}^{(\beta)}]
    &=
        \dfrac{\mu\ts{B}^2}{ (2\pi)^3 }
        \int_{\R^3} \dd^3 k
        \dfrac{ \kk^2 - k_3^2 }{ \kk (e^{\beta \kk} -1 ) } |\tilde \chi(\kk)|^2 \notag \\
        &\quad \times 
        \cos [\bk \cdot (\bx_\downarrow - \bx_\uparrow)]\,, \label{eq:nonlocal phase damp general Fourier} \\
    \mathcal{M}\ts{nl}^{(\beta)}
    &=
        \dfrac{-\mu\ts{B}^2}{ (2\pi)^3 }
        \int_{\R^3} \dd^3 k
        \dfrac{ \tilde \chi(\Omega - \kk) \tilde \chi(\Omega + \kk) }{ \kk (e^{\beta \kk} -1 ) } \notag \\
        &\quad \times 
        \text{Re}[ k_-^2 e^{ -\ii \bk \cdot (\bx_\downarrow - \bx_\uparrow) } ]\,,\label{eq:nonlocal amp damp general Fourier}
\end{align}
depend on the splitting configuration: \\
(i) When the spin is split along the $z$-axis: $\bx_\downarrow - \bx_\uparrow=[0,0,L]^\intercal$,
\begin{align}
    \text{Re}[\mathcal{D}\ts{nl}^{(\beta)}]
    &=
        \dfrac{2\mu\ts{B}^2 \sigma^2}{ \pi L^3 }
        \int_0^\infty \dd \kk 
        \dfrac{ e^{ -\kk^2 \sigma^2 } }{ e^{\beta \kk } -1 } \notag \\
        &\quad \times 
        [ -\kk L \cos (\kk L) + \sin (\kk L) ]\,, \\
    \mathcal{M}\ts{nl}^{(\beta)}&=0\,.
\end{align}
This means $\mathcal{M}\ts{nl}^{(\text{th})}=0$ because both the vacuum and thermal contributions are zero. \\
(ii) When the spin is split along the $x$-axis: $\bx_\downarrow - \bx_\uparrow=[L,0,0]^\intercal$,
\begin{align}
    \text{Re}[\mathcal{D}\ts{nl}^{(\beta)}]
    &=
        \dfrac{ \mu\ts{B}^2 \sigma^2 }{ \pi L^3 }
        \int_0^\infty \dd \kk 
        \dfrac{ e^{ -\kk^2 \sigma^2 } }{ e^{\beta \kk } -1 } \notag \\
        &\quad \times 
        [ \kk L \cos (\kk L) + (\kk^2 L^2 -1 )\sin (\kk L) ]\,,  \\
    \mathcal{M}\ts{nl}^{(\beta)}
    &=
        \dfrac{ -\mu\ts{B}^2 \sigma^2 e^{ -\Omega^2 \sigma^2 } }{ \pi L^3 }
        \int_0^\infty \dd \kk 
        \dfrac{ e^{ -\kk^2 \sigma^2 }}{ e^{\beta \kk} -1 } \notag \\
        &\quad\times 
        \Bigkagikako{
            3\kk L \cos (\kk L) + (\kk^2 L^2 -3) \sin (\kk L)
        }\,.
\end{align}

\begin{figure*}[tp]
\centering
\includegraphics[width=\linewidth]{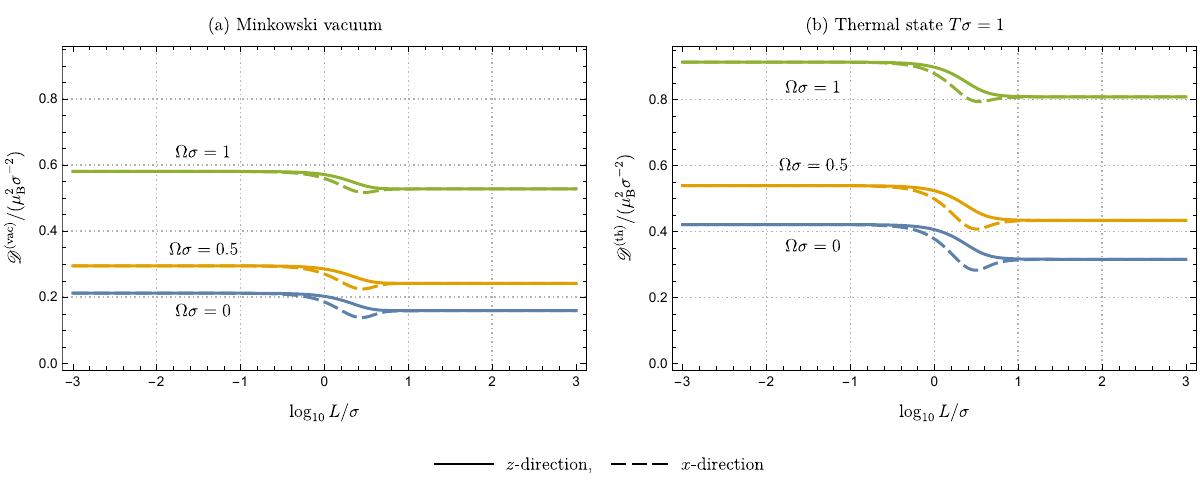}
\caption{Decoherence $\mathscr D/(\mu\ts{B}^2 \sigma^{-2})$ as a function of the splitting separation $L$ in units of typical interaction scale $\sigma$. 
The solid and dashed lines correspond to the splitting in the $z$- and $x$-directions, respectively. 
(a) When the field is initially prepared in the Minkowski vacuum; and (b) when the field is in the thermal state at temperature $T\sigma=1$. 
}
\label{fig:FigDecohEMBothvaryL}
\end{figure*}

\section{Decoherence of spin}\label{sec:Decoherence of spin}


\subsection{(De)coherence measure}\label{subsec:coherence measure}

We employ the coherence measure known as the \textit{$l_1$ norm of coherence} $\mathcal C_{l_1}$ \cite{Baumgratz.Coherence.2014, coherence.review.2017} defined as 
\begin{align}
    \mathcal C_{l_1}
    &\coloneqq
        \sum_{i \neq j} |\rho_{ij}|\,. 
\end{align}
We are particularly interested in a $4\times 4$ density matrix given in \eqref{eq:final density 4x4}. 
The corresponding $l_1$ norm of coherence is 
\begin{align}
    \mathcal C_{l_1}
    &=
        2 (|\rho_{14}| + |\rho_{23}|)\,.\notag
\end{align}
Based on \eqref{eq:elements of final state 4x4}, we have 
\begin{align}
    \mathcal C_{l_1}&=
        2|\rho_{14}^{(0)}|
        \Bigkako{
            1 - 
            \text{Re}
            [
                \mathcal{D}\ts{nl} + \mathcal{D}\ts{loc} + \mathcal{A}\ts{loc}
            ]
            + |\mathcal{M}\ts{nl}|
        } \notag \\
        &\quad
        + \mathcal{O}(\mu\ts{B}^4)\,, \label{eq:our coherence measure} 
\end{align}
where $\text{Re}[\mathcal{A}\ts{loc}]$ is identical to the ``average'' of transition probabilities: Eq.~\eqref{eq:real amplitude damp local}. 
From this expression, we define the measure for decoherence $\mathscr D$ as 
\begin{align}
    \mathscr D
    &\coloneqq
        \text{Re}
            [
                \mathcal{D}\ts{nl} + \mathcal{D}\ts{loc} + \mathcal{A}\ts{loc}
            ]
            - |\mathcal{M}\ts{nl}|~(\geq 0)\,, \label{eq:amount of decoherence}
\end{align}
which is independent of the initial amount of coherence. 
The explicit form for each term in \eqref{eq:amount of decoherence} is given in the previous section.

The local decoherences from amplitude damping $\text{Re}[\mathcal{A}\ts{loc}]$ and phase damping $\text{Re}[\mathcal{D}\ts{loc}]$ in $\mathscr D$ are always present regardless of the spin's superposition configuration. 
However, the nonlocal terms, $\text{Re}[\mathcal{D}\ts{nl}]$ and $|\mathcal{M}\ts{nl}|$, depend on the superposition configuration because they rely on the two-point correlation functions evaluated along the two trajectories, $\sx_\uparrow(\tau)$ and $\sx_\downarrow(\tau)$. 
The nonlocal term $|\mathcal{M}\ts{nl}|$ from amplitude damping is always nonnegative, and thus ``mitigates'' decoherence. 
In contrast, the nonlocal term $\text{Re}[\mathcal{D}\ts{nl}]$ from phase damping can be either positive or negative. 
Therefore, it can either worsen or reduce decoherence depending on the spatial superposition configuration and the chosen parameters.

In the following sections, we compute decoherence $\mathscr D$ divided by the dimensionless coupling constant $(\mu\ts{B}/\sigma)^2$.

\subsection{Splitting dependence}

In Fig.~\ref{fig:FigDecohEMBothvaryL}, we show the decoherence $\mathscr D/(\mu\ts{B}^2 \sigma^{-2})$ as a function of the logarithm of the splitting separation $L$ in units of the typical interaction time scale $\sigma$. 
The graph includes both the Minkowski vacuum case (a) and the thermal state case (at temperature $T\sigma = 1$) (b), each depicting $z$-splitting (solid lines) and $x$-splitting (dashed lines). 
From these figures, we can see that decoherence worsens as the energy gap increases due to amplitude damping, and higher temperatures cause more decoherence as expected.

Recall that the local decoherence terms are all independent of $L$. 
In Fig.~\ref{fig:FigDecohEMBothvaryL}, the values of the local decoherence correspond to those at large $L/\sigma$ because the nonlocal contributions vanish in the limit $L/\sigma \to \infty$. 
In contrast, decoherence $\mathscr D$ with smaller splitting separation $L$ includes both local and nonlocal contributions. 
Note that $\mathcal{M}\ts{nl} \to 0$ as $L \to 0$, thus the nonlocal phase damping $\mathcal{D}\ts{nl}$ accounts for the increase in decoherence near $L\approx 0$. 
The nonlocal amplitude damping $\mathcal{M}\ts{nl}$ mitigates decoherence, which is evident in the dashed lines around $\log_{10}L/\sigma \approx 0.5$. 
Note that the nonlocal terms depend on the splitting configuration because of the electromagnetic field's polarization. 
We compare this to the scalar field case in Appendix~\ref{app:UDW}.

\subsection{Temperature dependence}\label{subsec:temperature dependence}

\begin{figure}[t]
\centering
\includegraphics[width=\linewidth]{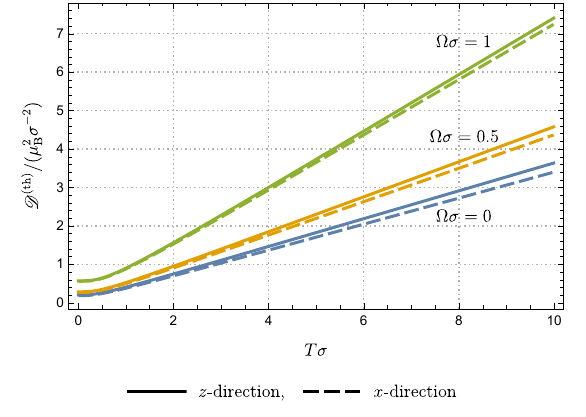}
\caption{Decoherence $\mathscr D/(\mu\ts{B}^2 \sigma^{-2})$ as a function of the field temperature $T\sigma$. 
The solid lines indicate splitting in the $z$-direction, and the dashed lines indicate splitting in the $x$-direction. 
Here, we have chosen $L/\sigma=1$. 
}
\label{fig:FigDecohEMTHBothvaryT}
\end{figure}

Here, we examine the temperature dependence of decoherence $\mathscr D$. 
In Fig.~\ref{fig:FigDecohEMTHBothvaryT}, we show decoherence $\mathscr D$ as a function of the field temperature $T$ in units of $\sigma$. 
We observe that decoherence increases monotonically with $T$ for all energy gaps $\Omega$.

Let us mathematically show that, for an arbitrary switching function $\chi(\tau)$, decoherence increases monotonically with temperature $\beta^{-1}$, based on the analysis in \cite{EduardoThermalMutual}. 
Consider two inverse temperatures $\beta\ts{H}$ and $\beta\ts{C}$ with $\beta\ts{H}^{-1}> \beta\ts{C}^{-1}$. 
Here, the subscripts `H' and `C' refer to hot and cold temperatures, respectively. 
We explicitly indicate the $\beta$-dependence as $\mathcal{D}\ts{nl}^{(\beta)}(\beta\ts{H})$. 
For convenience, we first show that the phase damping and amplitude damping terms satisfy
\begin{align}
    &\text{Re}[ \mathcal{D}\ts{loc}^{(\beta)}(\beta\ts{H}) ]
    +
    \mathcal{D}\ts{nl}^{(\beta)}(\beta\ts{H})
    \geq 
        \text{Re}[ \mathcal{D}\ts{loc}^{(\beta)}(\beta\ts{C}) ]
        +
        \mathcal{D}\ts{nl}^{(\beta)}(\beta\ts{C})\,, \label{eq:phase damp temp relations} \\
    &\text{Re}[ \mathcal{A}\ts{loc}^{(\beta)}(\beta\ts{H}) ]
    - 
    |\mathcal{M}\ts{nl}^{(\beta)}(\beta\ts{H}) |
    \geq 
        \text{Re}[ \mathcal{A}\ts{loc}^{(\beta)}(\beta\ts{C}) ]
        - 
        |\mathcal{M}\ts{nl}^{(\beta)}(\beta\ts{C})| \,,\label{eq:amp damp temp relations}
\end{align}
which lead to our conclusion 
\begin{align}
    \mathscr D^{\text{(th)}} (\beta\ts{H})\geq \mathscr D^{\text{(th)}} (\beta\ts{C})\,, \label{eq:temp and decoherence}
\end{align}
where $\mathscr D^{\text{(th)}}$ is understood as 
\begin{align}
    \mathscr D^{\text{(th)}}(\beta )
    &=
        \mathscr D^{\text{(vac)}}
        + 
        \mathscr D^{(\beta)}(\beta)\,.
\end{align}

First, let us focus on the phase damping terms and prove Eq.~\eqref{eq:phase damp temp relations}. 
From \eqref{eq:local phase damp general Fourier}, it is straightforward to show that 
\begin{align}
    &\text{Re}[ \mathcal{D}\ts{loc}^{(\beta)}(\beta\ts{H}) ] \notag \\
    &=
        \text{Re}[ \mathcal{D}\ts{loc}^{(\beta)}(\beta\ts{C}) ] 
        +
        \dfrac{\mu\ts{B}^2}{ (2\pi)^3 }
        \int_{\R^3} \dd^3 k
        \dfrac{\kk^2 - k_3^2}{\kk} |\tilde \chi(\kk)|^2 h(\kk)\,,
\end{align}
where 
\begin{align}
    h(\kk )
    &\coloneqq
        \dfrac{1}{ e^{\beta\ts{H} \kk} -1 } - \dfrac{1}{ e^{\beta\ts{C} \kk} -1 }\, (\geq 0)\,.
\end{align}
Likewise, the nonlocal phase damping term $\mathcal{D}\ts{nl}^{(\beta)}(\beta)$ in \eqref{eq:nonlocal phase damp general Fourier} results in  
\begin{align}
    \mathcal{D}\ts{nl}^{(\beta)}(\beta\ts{H})
    &=
        \mathcal{D}\ts{nl}^{(\beta)}(\beta\ts{C}) 
        + 
        \dfrac{\mu\ts{B}^2}{ (2\pi)^3 }
        \int_{\R^3} \dd^3 k
        \dfrac{\kk^2 - k_3^2}{\kk} \notag \\
        &\quad \times 
        |\tilde \chi(\kk)|^2 h(\kk) 
        \cos [\bk \cdot (\bx_\downarrow - \bx_\uparrow)] \notag \\
    &\geq 
        \mathcal{D}\ts{nl}^{(\beta)}(\beta\ts{C})  \notag \\
        &\quad 
        -
        \dfrac{\mu\ts{B}^2}{ (2\pi)^3 }
        \int_{\R^3} \dd^3 k
        \dfrac{\kk^2 - k_3^2}{\kk} 
        |\tilde \chi(\kk)|^2 h(\kk) \,,
\end{align}
and therefore, adding the two results yields Eq.~\eqref{eq:phase damp temp relations}.

We can also obtain Eq.~\eqref{eq:amp damp temp relations} through a similar calculation. 
Since $k_-^2=\kk^2 \sin^2 \theta e^{-\ii 2\varphi}$ in spherical coordinates $(\kk, \theta, \varphi)$, Eq.~\eqref{eq:nonlocal amp damp general Fourier} becomes 
\begin{align}
    &|\mathcal{M}\ts{nl}^{(\beta)}(\beta\ts{H})|
    \leq 
        |\mathcal{M}\ts{nl}^{(\beta)}(\beta\ts{C})| \notag \\
        &\quad
        +
        \dfrac{\mu\ts{B}^2}{ (2\pi)^3 }
        \int_{\R^3} \dd^3k\,
        \kk h(\kk) |\tilde \chi(\Omega -\kk) \tilde \chi(\Omega +\kk)|\,.
\end{align}
We also recall that $P_{\downarrow \to \uparrow}^{(\beta)}=P_{\uparrow \to \downarrow}^{(\beta)}$, and Eq.~\eqref{eq:transition prob thermal Fourier} gives us 
\begin{align}
    \text{Re}[ \mathcal{A}\ts{loc}^{(\beta)}(\beta) ]
    &=
        \dfrac{\mu\ts{B}^2}{2(2\pi)^3} 
        \int_{\R^3} \dfrac{ \dd^3 k }{ \kk( e^{\beta \kk } -1 ) }
        (\kk^2 + k_3^2) \notag \\
        &\quad \times 
        \kako{
            |\tilde \chi(\Omega + \kk)|^2
            +
            |\tilde \chi(\Omega - \kk)|^2
        }\,.
\end{align}
Since $\kk^2 + k_3^2=\kk^2(1+\cos^2\theta)$ in spherical coordinates, we obtain 
\begin{align}
    \text{Re}[ \mathcal{A}\ts{loc}^{(\beta)}(\beta\ts{H}) ]
    &\geq 
        \text{Re}[ \mathcal{A}\ts{loc}^{(\beta)}(\beta\ts{C}) ]
        +
        \dfrac{\mu\ts{B}^2}{2(2\pi)^3} 
        \int_{\R^3} \dd^3 k \, \kk h(\kk) \notag \\
        &\quad \times 
        \kako{
            |\tilde \chi(\Omega + \kk)|^2
            +
            |\tilde \chi(\Omega - \kk)|^2
        }\,.
\end{align}
Therefore, 
\begin{align}
    &\text{Re}[ \mathcal{A}\ts{loc}^{(\beta)}(\beta\ts{H}) ]
    - 
    |\mathcal{M}\ts{nl}^{(\beta)}(\beta\ts{H})| 
    \geq 
        \text{Re}[ \mathcal{A}\ts{loc}^{(\beta)}(\beta\ts{C}) ]
        - 
        |\mathcal{M}\ts{nl}^{(\beta)}(\beta\ts{C})| \notag \\
    &
    + 
    \dfrac{\mu\ts{B}^2}{2(2\pi)^3} 
    \int_{\R^3}\dd^3k\,
    \kk h(\kk) 
    \kako{
        |\tilde \chi(\Omega - \kk)|
        -
        |\tilde \chi(\Omega + \kk)|
    }^2 \notag \\
    &\geq 
        \text{Re}[ \mathcal{A}\ts{loc}^{(\beta)}(\beta\ts{C}) ]
        - 
        |\mathcal{M}\ts{nl}^{(\beta)}(\beta\ts{C})|\,,
\end{align}
which is Eq.~\eqref{eq:amp damp temp relations}. 
Hence, we have shown that the spatially superposed spin particle experiences monotonic decoherence with increasing temperature as described in Eq.~\eqref{eq:temp and decoherence}.

\subsection{Order estimation}
In this section, we bring back the fundamental constants $c, \hbar$, and $k\ts{B}$ and estimate the decoherence in a tabletop experiment. 
First, let us discuss the interaction time scale. 
While our Gaussian switching function is not exactly a compactly supported function, we use $10 \sigma$ as the effective total interaction time. 
This is because the range $\tau \in [ -5\sigma, 5\sigma ]$ is sufficient to provide accurate results, as shown by the normalized Gaussian integral: 
\begin{align}
    \int_{-5\sigma}^{5\sigma} \dd \tau\,
    \dfrac{1}{\sqrt{2\pi} \sigma} e^{ -\tau^2/2\sigma^2 }
    =
        0.999999\,. \notag
\end{align}
Thus, if the interaction time is 1\,s then $10\sigma = 1\,\text{s}$.

\subsubsection{Vacuum fluctuations}
Let us estimate the amount of decoherence caused by vacuum fluctuations. 
We can estimate the local decoherence terms $\text{Re}[\mathcal{A}\ts{loc}^{(\text{vac})}]$ and $\text{Re}[\mathcal{D}\ts{loc}^{(\text{vac})}]$ by expressing $\mu\ts{B}/\sigma$ and $\Omega \sigma$ in SI units [see Eqs.~\eqref{eq:transition prob Gaussian}, \eqref{eq:deexcitation transition prob Gaussian}, and \eqref{eq:vacuum local deco phase damp}]. 
Given that $\mu\ts{B}=e/(2m_e)$ in natural units, where $e$ and $m_e$ are the electron's charge and mass, we can use the fine-structure constant $\alpha\ts{f}=e/(4\pi)$ to convert $\mu\ts{B}/\sigma$ to SI units as follows. 
\begin{align}
    \dfrac{\mu\ts{B}}{\sigma}
    &=
        \dfrac{ \sqrt{ 4\pi \alpha\ts{f} } \hbar }{ 2m_e c^2 } \dfrac{1}{\sigma}
    \simeq
        1.2 \times 10^{-20}
        \kako{
            \dfrac{ 1\,\text{s} }{ 10 \sigma }
        }\,. 
\end{align}
From this value, the local phase damping decoherence term in \eqref{eq:vacuum local deco phase damp} can be estimated as follows. 
\begin{align}
    \text{Re}[\mathcal{D}\ts{loc}^{(\text{vac})}]
    &\simeq
        8.0 \times 10^{-42}
        \kako{
            \dfrac{ 1\,\text{s} }{ 10 \sigma }
        }^2\,. 
\end{align}
We recall that the energy gap in natural units is $\Omega = 2\mu\ts{B} B_0$, thereby 
\begin{align}
    \Omega \sigma 
    &=
        \dfrac{2 \mu\ts{B} B_0}{\hbar} \sigma
    \simeq 
        2.8 \times 10^{9} 
        \kako{
            \dfrac{B_0}{1\,\text{T}}
        }
        \kako{
            \dfrac{10\sigma}{ 1\,\text{s} }
        }\,.
\end{align}
Since this value is large, deexcitation primarily occurs. 
The local amplitude damping decoherence in \eqref{eq:real amplitude damp local}, using \eqref{eq:transition prob Gaussian} and \eqref{eq:deexcitation transition prob Gaussian}, can be approximated as 
\begin{align}
    \text{Re}[\mathcal{A}\ts{loc}^{(\text{vac})}]
    &\approx 
        \dfrac{ 1 }{3 \sqrt{\pi}}\dfrac{\mu\ts{B}^2}{\sigma^2} (\Omega \sigma)^3
        \quad
        ({\rm for } ~\Omega \sigma \gg 1) \notag \\
    &\simeq 
        6.2 \times 10^{-13}
        \kako{
            \dfrac{ 10 \sigma }{ 1\,\text{s} }
        }
        \kako{
            \dfrac{B_0}{1\,\text{T}}
        }^3\,.
\end{align}
Therefore, the dominant factor in local decoherence is amplitude damping.

We now estimate the nonlocal terms. 
For a splitting separation of $L\ll \sigma$ (e.g., $L\sim 100\,\mu\text{m}$), we saw in Eqs.~\eqref{eq:vac nonlocal phase damp x} and \eqref{eq:vac nonlocal phase damp z} that the nonlocal phase damping term $\text{Re}[\mathcal{D}\ts{nl}^{(\text{vac})}]$ is 
\begin{align}
    &\text{Re}[\mathcal{D}\ts{nl}^{(\text{vac})}]
    \approx 
        \dfrac{ \mu\ts{B}^2 }{ 6\pi \sigma^2 }
    = 
    8.0 \times 10^{-42}
        \kako{
            \dfrac{1\,\text{s}}{10\sigma}
        }^2
\,.
\end{align}
Although $\mathcal{D}\ts{nl}^{(\text{vac})}$ depends on the splitting configuration [see Eqs.~\eqref{eq:vacuum nonlocal dephasing z-split} and \eqref{eq:vacuum nonlocal dephasing x-split}], they both coincide with the local phase damping decoherence in this experimental setup. 
Moreover, the nonlocal mitigation term $|\mathcal{M}\ts{nl}^{(\text{vac})}|$ is exactly zero for the $z$-splitting [Eq.~\eqref{eq:Mnl z-split}], and $|\mathcal{M}\ts{nl}^{(\text{vac})}| \approx 0$ for the $x$-splitting [Eq.~\eqref{eq:Mnl x-split approx}].

Overall, local amplitude damping is the main source of decoherence from vacuum fluctuations, but it is still a small amount.

\subsubsection{Thermal noise}
Here, we estimate the decoherence from local amplitude damping caused by thermal noise \eqref{eq:beta amp damp local}. 
To this end, let us introduce
\begin{subequations}
\begin{align}
    \overline \mu\ts{B}
    &\coloneqq
        \dfrac{\sqrt{ 4\pi \alpha\ts{f} } k\ts{B}}{ 2 m_e c^2 }
    \simeq 
        2.6 \times 10^{-11} \,,\\
    \overline \Omega 
    &\coloneqq
        \dfrac{\hbar \Omega}{ k\ts{B}}
    =
        \dfrac{ 2\mu\ts{B} B_0 }{ k\ts{B} }
    \simeq 
        1.3
        \kako{
            \dfrac{B_0}{1\, \text{T}}
        }\,,\\
    \overline \sigma 
    &\coloneqq
        \dfrac{k\ts{B} \sigma}{ \hbar }
    \simeq 
        2.1 \times 10^9
        \kako{
            \dfrac{10\sigma}{ 1\,\text{s} }
        }\,.
\end{align}
\end{subequations}
We change the variable from $\kk$ to $r\equiv \beta \kk$ in \eqref{eq:beta amp damp local} and use the approximation $\cosh [2 (\overline \Omega/T) (\overline \sigma T)^2 r] \approx \exp[2 (\overline \Omega/T) (\overline \sigma T)^2 r]/2$ since $\overline \sigma$ is large. 
Then, changing the variable to $r'\equiv (r-\overline \Omega/T) \overline \sigma T$ and using $r'/(\overline \sigma T) \ll 1$, we get 
\begin{align}
    \text{Re}[\mathcal{A}\ts{loc}^{(\beta)}]
    &\approx 
        \dfrac{\overline \mu\ts{B}^2 \overline \sigma}{ 3\sqrt{\pi} } 
        \dfrac{\overline \Omega^3}{ e^{ \overline \Omega /T } -1 }
        [ 1 + \erf(\overline \Omega \overline \sigma) ] \,,
\end{align}
and at higher temperatures ($T \gg 1\,\text{K}$), 
\begin{align}
    \text{Re}[\mathcal{A}\ts{loc}^{(\beta)}]
    &\simeq 
            \dfrac{2\overline \mu\ts{B}^2 \overline \sigma}{ 3\sqrt{\pi} } 
              \overline \Omega^2T \notag\\
    &= 
        9.2 \times 10^{-11}
        \kako{
            \dfrac{10 \sigma}{1\,\text{s}}
        }
        \kako{
            \dfrac{ B_0 }{ 1\,\text{T} }
        }^2
        \kako{
            \dfrac{T}{ 100\,\text{K} }
        }\,.
\end{align}
Thus, local amplitude damping due to thermal noise is more significant than that caused by vacuum fluctuations.

\section{Conclusion}\label{sec:conclusion}

Maintaining coherence in a spin superposition state is crucial, particularly for generating a spatial superposition state of mass. 
However, the spin degree of freedom is inevitably influenced by the vacuum fluctuations of the electromagnetic field, leading to unavoidable decoherence. 
In this paper, we have investigated the decoherence of a spatially superposed spin particle caused by the spin-magnetic field coupling. 
To isolate the pure quantum field effect, we assumed that the spatially superposed spin trajectories are static (i.e., without acceleration and relative velocity) in both Minkowski vacuum and thermal states, and that the interaction occurs very smoothly.

The general expression for the coherence measure (the $l_1$ norm of coherence $\mathcal C_{l_1}$) in the leading order of the coupling constant is given by \eqref{eq:our coherence measure}, which is valid as long as the quantum field is initially in a quasifree state. 
We note that this expression can be straightforwardly generalized to an arbitrary globally hyperbolic spacetime. 
From this expression, we defined the amount of decoherence $\mathscr D$ in \eqref{eq:amount of decoherence}. 
Decoherence originates from amplitude damping and phase damping effects, and these come into $\mathscr D$ as local and nonlocal terms. 
The local terms correspond to the decoherence that solely depends on each branch of spatially superposed trajectories. 
In contrast, the nonlocal terms depend on the field correlation functions between the two branches, and it may mitigate decoherence depending on the configuration of the spatial superposition.

We then specified the field's state to the Minkowski vacuum and thermal state, and chose a Gaussian switching function to obtain the analytic form of $\mathscr D$. 
The result depends on how the spin's spatial superposition is configured: (i) when the spin's trajectory is split along the direction of the classical magnetic field in the spin's free Hamiltonian, and (ii) when it is split perpendicular to it. 
Such configuration dependence comes from the electromagnetic field's polarization, but it is invisible in a tabletop experiment because the separation distance is very small. 
We note that a qubit coupled to a quantum scalar field also exhibits such a nonlocal contribution, which is independent of the splitting configuration. 
Unlike the spin-$\frac{1}{2}$ particle case, the nonlocal term for a qubit mitigates the local decoherence when the splitting distance is small. 
See Appendix~\ref{app:UDW} for details.

We also analytically showed that a spatially superposed spin-$\frac12$ particle decoheres monotonically with the field's temperature. 
While the thermal contribution dominates over the decoherence due to vacuum fluctuations, the amount of decoherence is very small in a tabletop experiment.

Although our analysis focused on static superposition trajectories, it can be applied to any trajectory in spacetime. 
One interesting example is a uniformly accelerating trajectory. 
In this case, the spin particle experiences thermality due to the Unruh effect \cite{Unruh1979evaporation} since the pullback of the Wightman function along a single acceleration trajectory satisfies the KMS condition at the Unruh temperature $T\ts{U}=a/(2\pi)$, where $a$ is the proper acceleration. 
Consequently, any \textit{local} decoherence terms (i.e., $\text{Re}[\mathcal A\ts{loc}]$ and $\text{Re}[\mathcal D\ts{loc}]$) for the uniformly accelerating spin particle are equivalent to those for the static particle in a thermal bath. 
However, as pointed out in \cite{Danielson.BH.decohere.2022, Danielson.Killing.decohere.2023}, the uniformly accelerating spin particle is expected to decohere more by emitting soft photons in the long interaction limit, which does not occur in our thermal bath case. 
Such decoherence is characteristic when Killing horizons exist. 
We suspect that this soft photon-induced decoherence effect is encoded in the \textit{nonlocal} terms $\text{Re}[\mathcal D\ts{nl}]$ and $|\mathcal M\ts{nl}|$ in $\mathscr D$.

\acknowledgments{This work was supported by JSPS KAKENHI Grant Number 
JP23H01175. 
A.M. was supported by JSPS KAKENHI (Grants No. JP23K13103 and No. JP23H01175). 
}

\appendix
\section{Elements in coherence}\label{app:elements in coherence}

In this Appendix, we provide an explicit calculation for $\mathcal{M}\ts{nl}$ in \eqref{eq:Mnl z-split} and \eqref{eq:Mnl x-split}. 
The other nonlocal term $\text{Re}[\mathcal{D}\ts{nl}]$ can be obtained in a similar way.

The nonlocal term $\mathcal{M}\ts{nl}$ is defined as 
\begin{align}
    &\mathcal{M}\ts{nl}
    \coloneqq \notag \\
    &\mu\ts{B}^2
        \int_\R \dd \tau 
        \int_\R \dd \tau'\,
        \chi(\tau) 
        \chi(\tau')
        e^{ \ii \Omega (\tau + \tau') }
        \braket{\hat B_-(\sx_\uparrow) \hat B_-(\sx'_\downarrow)}_{\rho\ts{EM,0}}\,. \label{eq:Mnl}
\end{align}
Recall that the quantum magnetic field $\hatb{B}(\sx)$ can be expanded as [see Eqs.~\eqref{eq:B mode expansion} and \eqref{eq:magnetic mode function}]
\begin{align}
    \hatb{B}(\sx)
    &=
        \int_{\R^3} 
        \dd^3 k 
        \sum_{\lambda=1}^2 
        \kako{
            \bm{B}_{(\bk, \lambda)}(\sx) \hat a_{\bk, \lambda}
            + 
            \bm{B}_{(\bk, \lambda)}^*(\sx) \hat a_{\bk, \lambda}^\dag
        }\,,  \notag 
\end{align}
where 
\begin{align}
    \bm{B}_{(\bk, \lambda)}(\sx)
    &\equiv 
        \dfrac{ \ii \bk \times \bm{e}_{(\bk, \lambda)} }{ \sqrt{ (2\pi)^3 2 \kk } }
        e^{ -\ii \kk t + \ii \bk \cdot \bx }\,.\notag 
\end{align}
Denoting the components of the vector mode function $\bm{B}_{(\bk, \lambda)}(\sx)$ as 
\begin{align}
    \bm{B}_{(\bk, \lambda)}(\sx)=[ B_{(\bk, \lambda)}^{(1)}(\sx), B_{(\bk, \lambda)}^{(2)}(\sx), B_{(\bk, \lambda)}^{(3)}(\sx) ]^\intercal\,,
\end{align}
each component of the quantum magnetic field can be explicitly written as 
\begin{align}
    \hatb{B}(\sx)
    &=
        \begin{bmatrix}
            \hat B_1(\sx) \\
            \hat B_2(\sx) \\
            \hat B_3(\sx)
        \end{bmatrix} \notag \\
    &=
        \int_{\R^3} 
        \dd^3 k 
        \sum_{\lambda=1}^2 
        \Bigkako{
            \begin{bmatrix}
                B_{(\bk, \lambda)}^{(1)} \\
                B_{(\bk, \lambda)}^{(2)} \\
                B_{(\bk, \lambda)}^{(3)}
            \end{bmatrix}
            \hat a_{\bk, \lambda}
            + 
            \begin{bmatrix}
                B_{(\bk, \lambda)}^{(1)*} \\
                B_{(\bk, \lambda)}^{(2)*} \\
                B_{(\bk, \lambda)}^{(3)*}
            \end{bmatrix}
            \hat a_{\bk, \lambda}^\dag
        }\,, \notag 
\end{align}
where we have abbreviated the spacetime dependence in the second equation for simplicity.

Assuming that the field is initially in the Minkowski vacuum $\ket{0}$, the vacuum two-point correlation function of $\hat B_-(\equiv \hat B_1 - \ii \hat B_2)$ reads 
\begin{align}
    &\braket{0|\hat B_-(\sx_\uparrow) \hat B_-(\sx'_\downarrow)|0}
    =
        \int_{\R^3} 
        \dd^3 k 
        \sum_{\lambda=1}^2 \notag \\
        &\times 
        \kako{
                B_{(\bk, \lambda)}^{(1)} (\sx_\uparrow)
                - \ii
                B_{(\bk, \lambda)}^{(2)} (\sx_\uparrow)
            } 
        \kako{
                B_{(\bk, \lambda)}^{(1)*} (\sx_\downarrow')
                - \ii
                B_{(\bk, \lambda)}^{(2)*} (\sx_\downarrow')
            } \notag \\
    &=
        \int_{\R^3} 
        \dfrac{ \dd^3 k  }{ (2\pi)^3 2 \kk }
        (-k_-^2)
        e^{ -\ii \kk t_\uparrow + \ii \bk \cdot \bx_\uparrow }
        e^{ \ii \kk t_\downarrow - \ii \bk \cdot \bx_\downarrow }\,,
\end{align}
where we have used the identity 
\begin{align}
    \sum_\lambda
            (\bk \times \bm{e}_{(\bk, \lambda)})_- 
            (\bk \times \bm{e}_{(\bk, \lambda)}^*)_- 
        =
            -k_-^2\,,
\end{align}
where $(\bk \times \bm{e}_{(\bk, \lambda)})_- \equiv (\bk \times \bm{e}_{(\bk, \lambda)})_1 - \ii (\bk \times \bm{e}_{(\bk, \lambda)})_2$ and $k_- \equiv k_1 - \ii k_2$.

We now assume that each superposed trajectory is at rest in a single reference frame, so that $\sx_\uparrow$ and $\sx_\downarrow'$ can be written as $\sx_\uparrow(\tau)=(\tau, \bx_\uparrow)$ and $\sx_\downarrow(\tau')=(\tau', \bx_\downarrow)$. 
This implies that the proper spatial separation $L\equiv |\bx_\downarrow - \bx_\uparrow|$ is fixed throughout the interaction. 
The nonlocal term $\mathcal{M}\ts{nl}$ reduces to 
\begin{align}
    &\mathcal{M}\ts{nl}
    =
        \dfrac{\mu\ts{B}^2 }{ (2\pi)^3 2 } \notag \\
        &\quad \times 
        \int_{\R^3} \dd^3 k\,
        \dfrac{(-k_-^2)}{\kk} 
        e^{ -\ii \bk \cdot (\bx_\downarrow - \bx_\uparrow) }
        \tilde \chi(\Omega - \kk)
        \tilde \chi(\Omega + \kk)\,,  
\end{align}
where $\tilde \chi(\omega)$ is the Fourier transformation \eqref{eq:Fourier trans} of the switching function $\chi(\tau)$. 
We employ the spherical coordinates $(\kk, \theta, \varphi)$ for $\bk$ so that $\bk=[ \kk \sin \theta \cos \varphi, \kk \sin \theta \sin \varphi, \kk \cos \theta ]^\intercal$ and the integration measure reads $\dd^3 k=\kk^2 \sin \theta \dd \kk \dd \theta \dd \varphi$. 
Consequently the nonlocal term becomes 
\begin{align}
    &\mathcal{M}\ts{nl}
    =
    \dfrac{-\mu\ts{B}^2 }{ (2\pi)^3 2 } 
    \int_0^\infty \dd \kk\,
        \kk^3 \tilde \chi(\Omega - \kk)
        \tilde \chi(\Omega + \kk) \notag \\
    &\times 
        \int_0^\pi \dd \theta \,
        \sin^3 \theta 
        \int_0^{2\pi} \dd \varphi\,
        e^{-2\ii \varphi}
        e^{ -\ii \bk \cdot (\bx_\downarrow - \bx_\uparrow) }\,.
\end{align}
This expression for $\mathcal{M}\ts{nl}$ tells us that the result depends on the spin's splitting configuration $\bx_\downarrow - \bx_\uparrow$. 
In what follows, we examine two cases: (i) when the trajectory is split along the $z$-axis; and (ii) along the $x$-axis.

\noindent
(i) $z$-splitting: $\bx_\downarrow - \bx_\uparrow=[0, 0, L]^\intercal$

In this case, $\bk \cdot (\bx_\downarrow - \bx_\uparrow) = \kk L \cos \theta$, which is independent of $\varphi$. 
This means that the nonlocal term $\mathcal{M}\ts{nl}=0$ since 
\begin{align}
    \int_0^{2\pi} \dd \varphi\,
        e^{-2\ii \varphi}=0\,.
\end{align}

\noindent
(i) $x$-splitting: $\bx_\downarrow - \bx_\uparrow=[L, 0, 0]^\intercal$

Instead, if the spin trajectory is split along the $(x,y)$-plane, the inner product $\bk \cdot (\bx_\downarrow - \bx_\uparrow)$ depends on $\varphi$. 
In the case of $x$-splitting, we have 
\begin{align}
    \int_0^{2\pi} \dd \varphi\,
            e^{-2\ii \varphi}
            e^{ -\ii \kk L \sin \theta \cos \varphi }
    =-2\pi J_2(\kk L \sin \theta)\,,
\end{align}
where $J_2(z)$ is the Bessel function of the first kind. 
Further integrating over $\theta$ yields 
\begin{align}
    \mathcal{M}\ts{nl}
    &=
        \dfrac{-\mu\ts{B}^2 4\pi }{ (2\pi)^3 2 L^3 }
        \int_0^\infty \dd \kk\,
        \tilde \chi(\Omega - \kk)
        \tilde \chi(\Omega + \kk) \notag \\
        &\times 
        \Bigkagikako{
            3\kk L \cos (\kk L) + (\kk^2 L^2 -3) \sin (\kk L)
        }\,.
\end{align}
One obtains the final result \eqref{eq:Mnl x-split} by specifying the Gaussian switching function \eqref{eq:Gaussian switching}.


\section{UDW detector case}\label{app:UDW}
For a Unruh-DeWitt (UDW) detector, the interaction Hamiltonian only contains the amplitude damping part. 
In the interaction picture, it is given by 
\begin{align}
    \hat H\ts{I}^\tau(\tau)
    &=
        \lambda \chi(\tau ) (e^{\ii \Omega \tau} \hat \sigma_+ + e^{-\ii \Omega \tau} \hat \sigma_-)
        \otimes \hat \phi(\sx(\tau))\,. \label{eq:scalar interaction hamiltonian}
\end{align}
In $(3+1)$ dimensions, the coupling constant $\lambda$ is dimensionless. 
We assume the initial state 
\begin{align}
    \rho\ts{tot,0}
    &=
        \rho\ts{UDW,0} \otimes \rho_{\phi,0}\,,
\end{align}
where $\rho_{\phi,0}$ is a quasifree state for the scalar field and $\rho\ts{UDW,0}$ is the initial spatial superposition state of the detector corresponding to 
\begin{align}
        \alpha \ket{\uparrow, C_\uparrow} + e^{\ii \vartheta} \sqrt{1-\alpha^2} \ket{\downarrow, C_\downarrow}  \,.
\end{align}
The final density matrix $\rho\ts{UDW}$ of the detector reads 
\begin{align}
    \rho\ts{UDW}
    &=
        \begin{bmatrix}
            \rho_{11} & 0 &0 & \rho_{14} \\
            0 & \rho_{22} & \rho_{23} & 0 \\
            0 & \rho_{23}^* & \rho_{33} & 0 \\
            \rho_{14}^* & 0 &0 & \rho_{44} 
        \end{bmatrix}\,,
\end{align}
where 
\begin{subequations}
    \begin{align}
    \rho_{11}&=
        \rho_{11}^{(0)}[1  - P_{\uparrow \to \downarrow}^{C_\uparrow}(\Omega)]\,, \\
    \rho_{22}&=
        \rho_{44}^{(0)} P_{\downarrow \to \uparrow}^{C_\downarrow}(\Omega)\,, \\
    \rho_{33}&=
        \rho_{11}^{(0)} P_{\uparrow \to \downarrow}^{C_\uparrow}(\Omega)\,,\\
    \rho_{44}&=
        \rho_{44}^{(0)} [1 - P_{\downarrow \to \uparrow}^{C_\downarrow}(\Omega)]\,, \\
    \rho_{14}&=
        \rho_{14}^{(0)} ( 1 - \mathcal{A}\ts{loc} )\,,\\
    \rho_{23}&=
        \rho_{14}^{(0)*} \mathcal{M}\ts{nl}\,,
\end{align}
\end{subequations}
with 
\begin{subequations}
    \begin{align}
    \mathcal{M}\ts{nl}
    &\coloneqq 
        \lambda^2
        \int_\R \dd \tau 
        \int_\R \dd \tau'\,
        \chi(\tau) \chi(\tau') e^{ \ii \Omega (\tau+ \tau') } \notag \\
        &\quad\times 
        \braket{\hat \phi(\sx_\uparrow) \hat \phi(\sx_\downarrow')}_{\rho_{\phi,0}}\,,\\
    \mathcal{A}\ts{loc}
    &\coloneqq
        \lambda^2
        \int_\R \dd \tau 
        \int_\R \dd \tau'\,
        \Theta(t(\tau)  - t(\tau')) \notag \\
        &\quad\times 
        \chi(\tau) \chi(\tau') e^{ \ii \Omega (\tau - \tau') }
        \braket{\hat \phi(\sx_\uparrow) \hat \phi(\sx_\uparrow')}_{\rho_{\phi,0}} \notag \\
        &\quad
        +
        \lambda^2
        \int_\R \dd \tau 
        \int_\R \dd \tau'\,
        \Theta(t(\tau')  - t(\tau)) \notag \\
        &\quad\times 
        \chi(\tau) \chi(\tau') e^{ -\ii \Omega (\tau - \tau') }
        \braket{\hat \phi(\sx_\downarrow) \hat \phi(\sx_\downarrow')}_{\rho_{\phi,0}}\,.
\end{align}
\end{subequations}
Here, $\rho_{11}^{(0)}=\alpha^2$, $\rho_{14}^{(0)}=\alpha \sqrt{1-\alpha^2} e^{-\ii \vartheta}$, and $\rho_{44}^{(0)}=1 -\alpha^2$ [see Eq.~\eqref{eq:spin initial state}].

From the final density matrix, the $l_1$ norm of coherence can be computed as  
\begin{align}
    \mathcal C_{l_1}
    &=
        2 |\rho_{14}^{(0)}|
        (
            1 
            - 
            \text{Re}[
                \mathcal{A}\ts{loc}
            ]
            + |\mathcal{M}\ts{nl}|
        )\,,
\end{align}
and the amount of decoherence $\mathscr{D}$ is defined as 
\begin{align}
    \mathscr D
    &\coloneqq
        \text{Re}[
                \mathcal{A}\ts{loc}
            ]
            - |\mathcal{M}\ts{nl}|\,.
\end{align}
As in the case of the electromagnetic field \eqref{eq:real amplitude damp local}, we have 
\begin{align}
    \text{Re}[
                \mathcal{A}\ts{loc}
            ]
    &=
        \dfrac{1}{2} 
        [
            P_{\uparrow \to \downarrow}^{C_\uparrow}(\Omega)
            +
            P_{\downarrow \to \uparrow}^{C_\downarrow}(\Omega)
        ]\,.
\end{align}

\subsection{The Minkowski vacuum case}
Consider a static detector coupled to the massless quantum scalar field in the Minkowski vacuum in $(3+1)$-dimensional Minkowski spacetime. 
Employing the Gaussian switching function \eqref{eq:Gaussian switching}, the excitation transition probability can be computed as 
\begin{align}
    P_{\downarrow \to \uparrow}^{C_\downarrow}(\Omega)
    &=
        \dfrac{\lambda^2}{4\pi}
        \kagikako{
            e^{ -\Omega^2 \sigma^2 }
            - \sqrt{ \pi } \Omega \sigma 
            \erfc(\Omega \sigma)
        }\,.
\end{align}
Moreover, $\mathcal{M}\ts{nl}$ is independent of the direction of spatial split and it reads 
\begin{align}
    \mathcal{M}\ts{nl}
    &=
        \dfrac{\lambda^2 \sigma}{ 2\pi L } e^{-\Omega^2 \sigma^2} D^+(L/2\sigma)\,.
\end{align}
Note that the Dawson function $D^+(x)$ has this property 
\begin{align}
    \lim_{x\to 0} \dfrac{D^+(x)}{x}
    =
        1\,,
\end{align}
and therefore, $\mathcal{M}\ts{nl}$ takes the maximum value in the limit $L\to 0$: 
\begin{align}
    \lim_{L/(2\sigma) \to 0} \mathcal{M}\ts{nl}
    &=
        \dfrac{\lambda^2}{ 4\pi } e^{-\Omega^2 \sigma^2}\,.
\end{align}
This is a very different behavior compared to the electromagnetic field case, in which the nonlocal term $\mathcal{M}\ts{nl}$ vanishes at the coincidence limit $L\to 0$. 
Then, one can easily check that, in the coincidence limit $L\to 0$, a gapless detector ($\Omega=0$) does not decohere: $\mathscr D=0$. 
Note that if we start the interaction without splitting the trajectory, the density matrix for the UDW detector is a $2\times 2$ matrix. 
In this case, the decoherence measure is $\mathscr D= \text{Re}[\mathcal A\ts{loc}] - \text{Re}[\mathcal M\ts{nl}]$, which is different from the splitting scenario. 
However, since $\mathcal M\ts{nl}$ is real and positive, $\text{Re}[\mathcal M\ts{nl}]=|\mathcal M\ts{nl}|$, and the decoherence measure coincides.

\begin{figure*}[t]
\centering
\includegraphics[width=\linewidth]{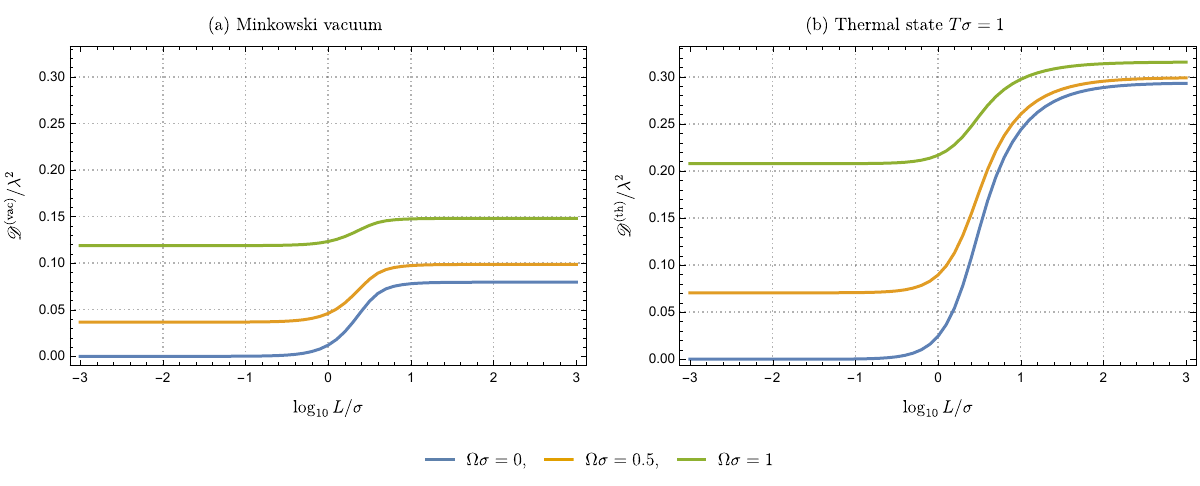}
\caption{The amount of decoherence $\mathscr D/\lambda^2$ as a function of $\log_{10} L/\sigma$ for $\Omega \sigma=0, 0.5$, and $1$ when 
(a) the scalar field is in the Minkowski vacuum state, and 
(b) in the thermal state at the temperature $T\sigma=1$. 
}
\label{fig:FigDecohScalarvaryLlarge}
\end{figure*}

On the other hand, the nonlocal term vanishes in the limit $L\to \infty$ and therefore
\begin{align}
    \lim_{L\to \infty} \mathscr D
    &=
        \text{Re}[\mathcal{A}\ts{loc}]\,. \notag
\end{align}
In contrast to the electromagnetic field case, decoherence reaches its maximum at large distances. 
See Fig.~\ref{fig:FigDecohScalarvaryLlarge}(a) for the separation dependence.

\subsection{The thermal state case}
As we saw in the electromagnetic field case, it is known that the thermal Wightman function $W\ts{th}(\sx, \sx')$ for a scalar field is a sum of the vacuum and the pure-thermal Wightman functions \cite{EduardoThermalMutual}: 
\begin{align}
    W\ts{th}(\sx , \sx')
    &=
        W\ts{vac}(\sx , \sx')
        +
        W_\beta(\sx , \sx')\,,
\end{align}
where 
\begin{align}
    W_\beta (\sx, \sx')
    &=
        \int_{\R^3} \dfrac{ \dd^3 k }{ (2\pi)^3 2\kk }
        \dfrac{ e^{ -\ii \kk (t-t')+ \ii \bk \cdot (\bx - \bx') } + \text{c.c.} }{ e^{\beta \kk } -1 }\,.
\end{align}
Therefore, the amount of decoherence for the thermal state $\mathscr D^{(\text{th})}$ can be decomposed as 
\begin{align}
    \mathscr D^{(\text{th})}
    &=
        \mathscr D^{(\text{vac})}
        +
        \mathscr D^{(\beta)} \,. \label{eq:thermal l1 norm of coherence}
\end{align}
where 
\begin{subequations}
\begin{align}
    \mathscr D^{(\text{vac})}
    &=
        \text{Re}[\mathcal{A}\ts{loc}^{(\text{vac})} ] - |\mathcal{M}\ts{nl}^{(\text{vac})}|\,, \\
    \mathscr D^{(\beta)}
    &=
        \text{Re}[\mathcal{A}\ts{loc}^{(\beta)} ] - |\mathcal{M}\ts{nl}^{(\beta)}|\,.
\end{align}
\end{subequations}
In this case, 
\begin{align}
    P_{\downarrow \to \uparrow}^{(\beta)}
    &=
        P_{\uparrow \to \downarrow}^{(\beta)} \notag \\
    &=
        \lambda^2 \int_{\R^3} \dfrac{\dd^3 k}{(2\pi)^3 2\kk} 
        \dfrac{ | \tilde \chi(\Omega + \kk) |^2
            +
            | \tilde \chi(\Omega - \kk) |^2 }{e^{\beta \kk} -1 }\,, \label{eq:scalar P beta general}\\
    \mathcal{M}\ts{nl}^{(\beta)} 
    &=
        2\lambda^2 
        \int_{\R^3} \dfrac{\dd^3 k}{(2\pi)^3 2\kk} 
        \dfrac{ \tilde \chi(\Omega + \kk) \tilde \chi(\Omega - \kk) }{e^{\beta \kk} -1 } \notag \\
        &\qquad \times 
        \cos [ \bk \cdot (\bx_\downarrow - \bx_\uparrow) ]\,.\label{eq:scalar M beta general}
\end{align}
In particular, if we choose the Gaussian switching function \eqref{eq:Gaussian switching}, these become 
\begin{align}
    P_{\downarrow \to \uparrow}^{(\beta)}
    &=
        \dfrac{\lambda^2 \sigma^2 }{ \pi } e^{ -\Omega^2 \sigma^2 }
        \int_0^\infty \dd \kk\,
        \dfrac{ \kk e^{ -\kk^2 \sigma^2 } }{ e^{ \beta \kk  } -1 } \cosh(2 \Omega \kk \sigma^2)\,,\\
    \mathcal{M}\ts{nl}^{(\beta)}
    &=
        \dfrac{ \lambda^2 \sigma^2 }{ \pi L } e^{ -\Omega^2 \sigma^2 }
        \int_0^\infty \dd \kk\,
        \dfrac{ e^{ -\kk^2 \sigma^2 } }{ e^{ \beta \kk  } -1 } \sin(\kk L)\,.
\end{align}
The $L$-dependence is depicted in Fig.~\ref{fig:FigDecohScalarvaryLlarge}(b) when the field temperature is $T\sigma=1$.

Following the same proof in Sec.~\ref{subsec:temperature dependence}, we can show that for an arbitrary switching function $\chi(\tau)$, decoherence monotonically increases with temperature $\beta^{-1}$. 
That is, for $\beta\ts{H}^{-1}> \beta\ts{C}^{-1}$, 
\begin{align}
    \mathscr D^{(\text{th})}(\beta\ts{H}) \geq \mathscr D^{(\text{th})} (\beta\ts{C})\,.
\end{align}

A recent paper \cite{wilsongerow2024decoherencewarmhorizons} discusses the decoherence of a uniformly accelerating UDW detector. 
Let us compare our UDW setup with theirs. 
In our study, we consider the qubit, the COM, and the field degrees of freedom: $\hil\ts{UDW} \otimes \hil\ts{COM}\otimes \hil_\phi$. 
The COM degree of freedom is assumed to be fixed after splitting, so it does not decohere. 
Specifically, the interaction Hamiltonian in the Schr\"odinger picture takes the form $\hat \sigma_1 \otimes \id\ts{COM} \otimes \hat \phi$, where $\hat \sigma_1 \in \mathcal B(\hil\ts{UDW})$. 
Thus, this Hamiltonian describes the decoherence of a UDW detector (spin degree of freedom) caused by amplitude damping. 
In contrast, \cite{wilsongerow2024decoherencewarmhorizons} considers $\hil\ts{COM} \otimes \hil_\phi$ and examines the COM's decoherence caused by the quantum field. 
Similar to our case, each wavepacket of the superposed system is assumed to be narrow and localized, making $\hil\ts{COM}$ a two-dimensional Hilbert space. 
Hence, they treat the COM degree of freedom as a UDW detector. 
The interaction Hamiltonian is derived from $\ket{C_\uparrow}\bra{C_\uparrow} \otimes \hat \phi(\sx_\uparrow) + \ket{C_\downarrow}\bra{C_\downarrow} \otimes \hat \phi(\sx_\downarrow)$, which can be reduced to $\hat \sigma_3 \otimes \varepsilon^\mu \partial_\mu \hat \phi(\sx)$, where $\hat \sigma_3 \in \mathcal{B}(\hil\ts{COM})$ and $\varepsilon^\mu$ describes the displacement between the two trajectories. 
This describes the decoherence of the COM degree of freedom caused by phase damping.

\bibliography{ref}

\end{document}